\documentstyle[11pt,aaspp4]{article}
\begin{document}
\newcommand{\lya}{Lyman~$\alpha$}
\newcommand{\lyb}{Lyman~$\beta$}
\newcommand{\za}{$z_{\rm abs}$}
\newcommand{\ze}{$z_{\rm em}$}
\newcommand{\cmtwo}{cm$^{-2}$}
\newcommand{\nhi}{$N$(H$^0$)}
\newcommand{\degpoint}{\mbox{$^\circ\mskip-7.0mu.\,$}}
\newcommand{\kms}{\,km~s$^{-1}$}      
\newcommand{\minpoint}{\mbox{$'\mskip-4.7mu.\mskip0.8mu$}}
\newcommand{\peryr}{\mbox{$\>\rm yr^{-1}$}}
\newcommand{\secpoint}{\mbox{$''\mskip-7.6mu.\,$}}
\newcommand{\sqdeg}{\mbox{${\rm deg}^2$}}
\newcommand{\squig}{\sim\!\!}
\newcommand{\subsun}{\mbox{$_{\twelvesy\odot}$}}
\newcommand{\et}{{\it et al.}~}

\def\ltsima{$\; \buildrel < \over \sim \;$}
\def\simlt{\lower.5ex\hbox{\ltsima}}
\def\gtsima{$\; \buildrel > \over \sim \;$}
\def\simgt{\lower.5ex\hbox{\gtsima}}
\def\arcs{$''~$}
\def\arcm{$'~$}
\def\erf{\mathop{\rm erf}}
\def\erfc{\mathop{\rm erfc}}
\title{LYMAN ALPHA IMAGING OF A PROTO--CLUSTER REGION AT ${\rm \bf \langle z \rangle =3.09}$
\altaffilmark{1}}
\author{\sc Charles C. Steidel\altaffilmark{2,3}, Kurt L. Adelberger, Alice E. Shapley}
\affil{Palomar Observatory, California Institute of Technology, MS 105--24, Pasadena, CA 91125}
\author{\sc Max Pettini}
\affil{Institute of Astronomy, University of Cambridge, Madingley Road, Cambridge CB3 0HA, UK}
\author{\sc Mark Dickinson and Mauro Giavalisco}
\affil{Space Telescope Science Institute, 3700 San Martin Drive, Baltimore, MD 21218}

\altaffiltext{1}{Based on data obtained at the Palomar Observatory 
and the W.M. Keck Observatory. The W. M. Keck Observatory
is operated as a scientific partnership among the California Institute of Technology, the
University of California, and NASA, and was was made possible by the generous financial
support of the W.M. Keck Foundation.
} 
\altaffiltext{2}{NSF Young Investigator}
\altaffiltext{3}{Packard Fellow}
\begin{abstract}
We present very deep narrow band observations of a volume containing a
significant over-density of galaxies 
previously discovered in our survey for continuum--selected Lyman break galaxies at
redshifts $2.7 \simlt z \simlt 3.4$. The new observations are used
in conjunction with our spectroscopic results on LBGs 
to compare the effectiveness of continuum and emission line searches for star forming galaxies
at high redshift, and to attempt to extend the search for members of the
structure at $\langle z \rangle= 3.09$ to much fainter continuum luminosities. 
The 8\minpoint7 by 8\minpoint9 
field contains a very high surface density of emission line candidates,
approximately 6 times higher than in published 
blind narrow--band searches to comparable
depth and at similar redshift. This level of density enhancement for the $\langle z \rangle =3.09$
structure is consistent with that inferred from the analysis of the
spectroscopic Lyman break galaxy sample in the same region ($6.0\pm 1.2$), but extends to continuum
luminosities up to 2 magnitudes fainter. 
We find that only $\sim 20-25\%$ of all galaxies
at a given UV continuum luminosity would be flagged as narrow--band excess objects subject
to the typical limits $W_{\lambda} > 80$ \AA\ in the observed frame. 
The remainder have lines that are too weak ($<20$ \AA\ rest equivalent width) 
to make selection by narrow--band excess effective. 
There is no evidence for a significantly higher fraction of
large Lyman $\alpha$ line equivalent widths at faint continuum luminosities.  

We have also discovered two extremely bright, large, and diffuse Lyman $\alpha$ emitting ``blobs'', 
which are apparently associated with, but not centered on, previously known Lyman break galaxies at the
redshift of the $\langle z \rangle =3.09$ structure. These nebulae have physical extents
$\simgt 100$h$^{-1}$ kpc and Lyman $\alpha$ line fluxes of $\sim 10^{-15}$
ergs s$^{-1}$ cm$^{-2}$, both factors of $\sim 20-40$ times larger than the typical
line emitters at the same redshifts in the field. In many respects these ``blobs''
resemble the giant Lyman $\alpha$ nebulae associated with high redshift radio galaxies,
but have $<$1\% of the associated radio continuum flux and no obvious
source of UV photons bright enough to excite the nebular emission. While the nature of
these blobs remains unclear, it is possible that they are excited
by continuum sources that are heavily obscured along our line of sight, or that they
are associated with cooling--flow--like phenomena.

The effectiveness of narrow--band imaging for isolating large, albeit incomplete,
samples of high redshift galaxies over a broad range of continuum luminosity 
makes the technique particularly well--suited to ``mapping''
known or suspected structures at high redshift. By combining the 24 spectroscopic
members of the $z=3.09$ ``spike'' with the narrow--band candidates, we are able
to produce a sample of 162 objects in a single relatively small field which 
are either known or likely members of this large structure.  A smoothed surface density
map shows 3 regions exceeding a local
over-density of $\delta \rho/\rho=1$ on $1h^{-1}$ Mpc (co--moving) scales; interestingly,
one is centered on a $z=3.083$ QSO, and another on one of the giant Lyman $\alpha$ nebulae.
There is thus circumstantial evidence that the Lyman $\alpha$ nebulae may, like QSOs,
be linked to the sites of the largest density enhancements at high redshifts.

\end{abstract}
\newpage

\section{INTRODUCTION}

Searches for high redshift galaxies using the Lyman $\alpha$ emission line have
long held the promise of isolating distant galaxies undergoing strong
episodes of star formation.
There have now been many cases in which ``targeted''
narrow--band searches in the vicinity of known objects (QSOs, radio galaxies, QSO
absorption systems, etc.) have successfully yielded companion or associated objects by virtue
of strong Lyman alpha emission found with the tuned narrow filter (e.g., Djorgovski \et 1985, Lowenthal \et
1991, Francis \et 1996, Le F\'evre \et 1996, Hu \& McMahon 1996, 
Pascarelle \et 1996, Moller \& Warren 1998). 
However,  until very recently, essentially all ``blank field''
or ``blind'' searches for Lyman alpha emitting objects have had null results.
It has been recognized for more than 20 years (cf. Meier 1976) that the nature of Lyman alpha
as a resonance line makes it particularly susceptible to selective extinction by small amounts
of dust in young galaxies, and indeed a series of theoretical papers in the early 1990's
(Charlot \& Fall 1991, 1993; Chen \& Neufeld 1994),
and the generally null observational results available at the time, pushed the high redshift
galaxy searches in other directions. One alternative was to exercise the same principles as
for the Lyman alpha emission line searches, but image in the near--IR (rest--frame
optical) where familiar nebular lines like H$\alpha$, which are expected to be much
less affected by dust and radiative transfer effects, are accessible (cf. 
Thompson \et 1996, Malkan \et 1996, Teplitz \et 1998, Mannucci \et 1998). Most
recently, slitless spectroscopy with NICMOS on board the {\it Hubble Space Telescope} has 
been used successfully
to find H$\alpha$ emitting galaxies at high redshifts (Yan \et 1999). Another approach has 
been to use the Lyman continuum photoelectric absorption
feature at 912 \AA\ and broad--band photometry to isolate larger
swaths of redshift space at high redshift (e.g., 
Steidel, Pettini, \& Hamilton 1995, Steidel \et
1996, 1999; Madau \et 1996, Lowenthal \et 1997). All of these techniques have
enjoyed success in the past few years. The Lyman break method in particular has allowed
fairly extensive ``blind'' surveys for high redshift galaxies, whose resulting large
samples introduce, among other things,  the new possibility of studying the large--scale distribution
of galaxies when the universe was only 10--15\% of its current age
(Steidel \et 1998; Giavalisco \et 1998; Adelberger \et 1998). A combination
of extensive ground--based photometric and spectroscopic,  and {\it HST} photometric
Lyman break galaxy (LBG) surveys has also allowed a census of the distribution of
UV luminosity in galaxies at $z \sim 3$ 
and estimates of the total
star formation and metal production rate in galaxies at $2 \simlt z \simlt 4.6$
(Madau \et 1996, 1998; Steidel \et 1999). 

However, the Lyman break method requires very deep broad--band images in order to discern
significant continuum breaks in the spectral energy distribution of faint galaxies, and
this limits the method to continuum magnitudes ${\cal R} \simlt 25-26$ from the ground.
This method also becomes increasingly more difficult beyond $z \sim 4$, and essentially
impossible from the ground beyond $z \sim 6$. These limitations---and the desire
to go to still higher redshifts, and to fainter continuum luminosities--- make re-visiting
the Lyman $\alpha$ emission line search technique attractive.
Within the past year, Cowie \& Hu (1998; CH98) and Hu, Cowie, \& McMahon (1998) have 
demonstrated that blank field Lyman $\alpha$ searches can be successful provided
that they go only a few times deeper than the many previous attempts. They also
demonstrated that some objects could be discovered, using narrow--band imaging,
that are too faint in the continuum to have been found using  
techniques based on ground--based continuum imaging. This suggests that Lyman alpha imaging
might be used successfully to push surveys down to very faint continuum luminosities,
complementing the Lyman break technique's efficient selection of relatively bright objects.

Unfortunately, Lyman $\alpha$ selection is subject to serious problems of its own. 
Aside from the obvious limitation that survey volumes are small 
due to the typically narrow filters used to maximize the contrast of
line emission versus continuum,
it is also very difficult to extract quantitative information about the high
redshift galaxy population based on Lyman alpha line strength (flux or equivalent width) alone.
A substantial fraction of continuum--selected
high redshift galaxies do not have significant Lyman alpha emission (Steidel \et 1996,1998; Lowenthal
\et 1997),
and indeed, as we will discuss, about half of the relatively bright LBGs in the redshift interval
$z=2.5-4.5$ show strong Lyman $\alpha$ in {\it absorption},
due to complete scattering of Lyman alpha photons from our line of sight
by optically thick H~I with a large velocity spread. 
While there has been some (largely anecdotal) indication that,
as one goes to fainter intrinsic luminosities, the propensity to
exhibit strong Lyman alpha emission may increase, this has
not yet been properly quantified. Very few
of the objects found so far from narrow--band searches at any
continuum flux level have Lyman alpha rest equivalent widths approaching
that expected from simple case-B recombination and excitation by 
stellar UV radiation field, $\sim 150$ \AA\ (cf. Charlot \& Fall 1993), 
and in general Lyman $\alpha$ line luminosity is only weakly correlated with
continuum luminosity.  
CH98 have argued that the surface density per unit redshift of strong Lyman alpha emitters is
about the same as that of objects found using continuum breaks, but as we
show below, this is not a meaningful comparison in any quantitative
sense, since the LBG samples are continuum flux limited and the strong
line emitters are drawn from a huge range of continuum luminosity. A more
quantitative statement can be made 
by comparing the UV--continuum luminosity density of Lyman alpha emitters
to that obtained by integrating over the $z \sim 3$ LBG luminosity function 
of Steidel \et (1999): the continuum selected objects to ${\cal R}\sim 27$
produce roughly 4 times more integrated star formation per unit co--moving
volume than the Lyman $\alpha$ selected sample of
CH98, without applying extinction correction to either sample. The factor is
even larger if the  
Lyman alpha line luminosities, rather than the
UV continua, are used to estimate the integrated star formation rate of the
narrow-band-selected
population. The luminosity functions of emission--line and continuum selected
objects will be compared in more detail in \S 4 below.

Steidel \et (1998) presented spectroscopic evidence for a large
structure of galaxies at $z=3.09\pm0.03$ based on initial results from
a survey for $z \sim 3$  LBGs in the SSA22 survey field; 
since the time of that paper, the spectroscopic
sample in the SSA22 field has been increased by approximately 50\% and it has
been demonstrated that such large structures are not uncommon
at $z \sim3$ (Adelberger \et 1998). In view of the interest of
both cross--correlating samples which one obtains from the LBG
continuum selection and from narrow--band Lyman alpha imaging,
and attempting to assess the number of intrinsically faint
galaxies associated with this large over--density, in the spring
of 1997 we purchased a custom narrow--band filter 
tuned to the redshift (and redshift interval) of Lyman $\alpha$ for galaxies in the
structure in the SSA22a,b field. 

The current redshift histogram for the northern half of the 9\arcm\ by
18\arcm the SSA22 field (dubbed ``SSA22a'' in Steidel \et 1998), is 
shown in Figure 1. The binning has been chosen so that the bin width
corresponds to the range in redshift spanned by a filter with a FWHM of
80\AA\ for Lyman $\alpha$ near $z \sim 3$, with the phase of the bins
chosen so that there is a bin centered on $z=3.09$, the median redshift
of the structure identified by Steidel \et 1998. The selection function
shown in Figure 1 is based on the more recent results from the full 
survey for $z \sim 3$ Lyman break galaxies covering
0.3 square degrees (Steidel \et 1999, Adelberger \et 2000) and reflects
the fact that the SSA22a field has a $\sim$30\% higher space density
than an average field of the same volume\footnote{i.e., the ratio of each bin in the
observed histogram
to the selection function reflects the factor by which 
the volume density in the bin is enhanced with respect to an average volume of the
same size at the same redshift. Thus, the integrated area under the selection
function is smaller by $\sim 30$\% than the observed number of galaxies in the
SSA22a field.}. Thus, in the volume probed by our narrow band imaging (see below), 
the space density of LBG to our magnitude limit of ${\cal R}=25.5$ is $6.0\pm1.2$
times higher than average. The implied over-density
of $5.0 \pm 1.2$ is slightly larger than that estimated by Steidel \et 1998.  

In this paper, we report on the results of very deep narrow--band imaging
in the Lyman $\alpha$ line at the redshift of the known large structure, likely
the progenitor of a present--day rich cluster of galaxies (cf. Steidel \et 1998,
Governato \et 1998). We compare
the results to those obtained in ``blank'' fields, and we inter-compare
the galaxy samples selected using the LBG technique and narrow--band
selection. A new phenomenon of very luminous, but diffuse, Lyman
alpha nebulae is presented, and we discuss the possible nature of these objects. 
Finally, we discuss the virtues and limitations of the narrow--band technique, and
other methods that depend on strong emission lines, for
assessing the galaxy populations at high redshift.

\section{OBSERVATIONS}

All of the imaging data were obtained using the
COSMIC prime focus camera (Kells \et 1998) on the Palomar 200--inch (5.08m) telescope.  
The detector used is a thinned, AR--coated Tektronix 2048 x 2048
CCD, providing a plate scale of 0.283 arcsec pixel$^{-1}$ and a field
of view of 9\minpoint7 by 9\minpoint7 on the sky. 
Most of the broad--band images used here had been obtained during the
years 1995-97 as part of our survey for Lyman break galaxies; the only
new broad--band data include a 2.75 hour V band image, which is combined
with our existing G band image (4730/1100) to create the continuum image
as described below. The narrow--band images were obtained in 1997 September
and 1998 August, September, and October\footnote{Essentially all of the useful
data were obtained in 1998, due to poor weather in 1997.} using a specially
designed interference filter with a bandpass centered at 4970 \AA\ (corresponding
to Lyman $\alpha$ at $z=3.09$) and a bandwidth of FWHM=80\AA. The data
were reduced using standard procedures, although only dome and twilight sky flats (as opposed to
flat fields produced from dis-registered dark sky images) were
used for the narrow--band images in order to prevent the suppression of
extended, diffuse emission. The final
narrow--band image consists of a total of 15.8 hours of integration under
photometric conditions; the seeing in the stacked image was 1\secpoint2 FWHM.
Both the broad--band data and narrow--band data were calibrated onto 
the AB system using spectrophotometric standard stars from the list of Massey \et (1988).
All magnitudes and colors quoted in this paper are on the AB system, and are corrected
for Galactic extinction of E(B-V)=0.08 based upon IRAS 100$\mu$m cirrus emission maps (cf. 
Schlegel, Finkbeiner, \& Davis 1998). This extinction correction results in similar
mean colors for the faint galaxies in the SSA22 field and in other, less reddened, fields
in our survey.
Note that this extinction correction is somewhat
larger than that assumed by CH98, and would result in a relative offset near 
5000 \AA\ of $\sim 0.11$ magnitudes in the sense that our (corrected) magnitudes would
be brighter.  

The narrow--band image is among the deepest ever obtained, achieving
a 1 $\sigma$ surface brightness limit within an aperture the size of the
seeing disk of ${\rm NB=28.53}$ mag, or $\simeq 27.5$ in a 3\arcs\
diameter aperture, comparable to that achieved in the deepest narrow--band 
exposures (obtained using the Keck telescopes) presented by CH98. 

With an effective wavelength of 4730 \AA, the G filter is slightly bluer than
the 4970\AA\ narrow band filter. In order to
more precisely sample the continuum at the same effective wavelength as
the narrow band filter, and to make the continuum image as deep as possible
for reasons that will become clear below, we constructed an image that
we will refer to as the ``GV continuum'' using a linear combination
(GV$={\rm 2G+V}$) of
the deep G and V images after scaling them to the same photometric
zero point. Because the net image quality in the GV continuum image was somewhat
better than in the narrow--band image, we slightly smoothed it
to match the FWHM=1\secpoint2 image quality. The GV continuum image, which
has an effective exposure time of $\sim 6$ hours, reaches
a 1 $\sigma$ photometric limit of ${\rm GV}\simeq 27.9$ in a 3\arcs\ diameter aperture.    

The final continuum and narrow--band images were trimmed and brought into
registration with the images we have used previously for our LBG
survey. The total size of the field is 8\minpoint74 by 8\minpoint89, for
a total solid angle of 78 arcmin$^{2}$. The effective volume probed by
the narrow--band imaging has (co--moving) transverse dimensions of 7.7h$^{-1}\times
7.8$h$^{-1}$ Mpc, and the half--power points of the filter correspond to a
co--moving depth along the line of sight of 23$h^{-1}$ Mpc and a total volume
of 1380h$^{-3}$ Mpc$^{3}$, assuming an Einstein-de Sitter cosmology. 
The corresponding dimensions 
for a cosmology with $\Omega_m=0.3$, $\Omega_{\Lambda}$=0.7 are 
11.5h$^{-1}$ $\times 11.7$h$^{-1} \times$43h$^{-1}$ Mpc. 

Several photometric catalogs were produced from the data, using 
methods very similar to those described in Steidel, Pettini, \& Hamilton (1995). First, a catalog in
which object detection was performed on the NB image was created. Object detection
was performed using a modified version of FOCAS (Jarvis \& Tyson 1982), and was
based on thresholding (at 3$\sigma$ per pixel above local sky) 
a slightly smoothed image subject to the criterion
that the area of detected objects must be greater than or equal to the size of the seeing disk
FWHM. FOCAS ``total'' magnitudes are calculated by growing the isophotal apertures 
such that the aperture is doubled in area.  The FOCAS isophotal apertures are then
applied to the GV continuum image to calculate colors; thus, the colors are calculated
using apertures defined in the NB image. A similar catalog was generated using
detections in the GV continuum image, and applying the isophotes to the NB image-- the
purpose of this second catalog is to evaluate the number of objects in the field
which have significant narrow--band {\it deficits}, which several of the known
objects in the $z=3.09$ structure are expected to exhibit because their spectra
show very strong Lyman $\alpha$ in absorption (rather than emission). 
Finally, a third catalog was generated by performing object detection on a
continuum--subtracted version of the NB image, and on which both significant
NB excess and NB deficit objects were retained. The motivation for this last 
catalog is that it is the most straightforward for measuring the total line
flux from narrow--band excess objects, and provides a sanity check for
the reality of identified objects in the first two catalogs. It also better allows
for the fact that in a number of cases the morphology and/or positions of objects
with significant line emission are not identical to those in the continuum image,
so that colors in matched isophotal apertures would not necessarily produce
accurate line fluxes.  

In all cases, when the flux in an isophotal aperture was smaller than $+1\sigma$
for the sky fluctuations expected in the aperture, an upper limit of $+1\sigma$
was assigned for the measured flux.

\section{NARROW--BAND EXCESS OBJECTS}

A total of 2919 objects with $NB \le 25.5$ were retained in the NB-selected 
catalog; 2089 of these have $NB \le 25.0$. The results for the colors of
all of the objects in the catalog are shown in Figure 2. Because there
is significant intrinsic scatter in the ${\rm GV-NB}$ color for general field
galaxies, the significance of departures from zero color for the
assessment of emission line objects should be judged relative to the
locus defined by the bulk of the objects at a given apparent magnitude.
This scatter is generally several times larger than the formal
photometric uncertainties, until the faintest magnitudes are reached.

The effective widths of the GV and NB passbands are 1600\AA\ and 80\AA, respectively.
An emission line with an equivalent width of 80\AA\ would double the count
rate in the NB relative to a pure continuum source, resulting
in a color ${\rm GV-NB}=0.7$.
Following CH98 for consistency, the dotted line in
Figure 2 indicates this ${\rm GV-NB}=0.7$ limit. 
It is clear that the photometric scatter begins to place a significant
number of objects above this threshold for $NB > 25$; thus, although the NB
photometric completeness should be reasonable to $NB \sim 25.5$, for the
purposes of statistical samples we will conservatively truncate the catalog
at $NB=25.0$. This also allows direct comparison with the largest ``field''
sample compiled by CH98 from a combination of SSA22 and HDF
narrow band fields obtained at $z=3.4$ with a FWHM=77\AA\ filter. 
An additional, larger, sample of NB excess objects to ${\rm NB =25.5}$, has
been retained for some purposes (see \S 6); this sample has a more conservative
color cut of ${\rm GV-NB >1}$ for ${\rm 25 \le NB < 25.5}$.\footnote{It is worth
mentioning the (perhaps obvious) point that narrow--band catalogs defined in this way 
are neither NB flux limited nor line equivalent width limited for objects with
a range of continuum luminosity. For example, a NB--selected sample is complete
with respect to continuum luminosity and line equivalent width  
for continuum luminosities ${\rm GV < NB_{lim}+C_{lim}}$, where 
${\rm NB_{lim}}$ is the magnitude limit of the narrow band catalog, and ${\rm C_{lim}}$
is the adopted ${\rm GV-NB}$ contrast threshold.  However, fainter continuum objects will be included
only if they satisfy
${\rm GV-NB \ge GV-NB_{lim}}$, where ${\rm GV -NB_{lim} > C_{lim}}$, so that
fainter objects need larger equivalent widths to be included in the sample.
Similarly, the samples are also incomplete with respect to emission line flux, since continuum
objects with the same line flux will have line equivalent widths (i.e., narrow-band/continuum
contrast) that decrease with increasing continuum luminosity. We address these points
further in \S 4.} Table 1 contains a summary of the various samples discussed in
this section.

There is a total of 77 objects whose colors indicate significant narrow--band
excess as defined using the $NB \le 25$, $W_{\lambda} \ge 80$ \AA\ criteria. 
As discussed by CH98 and others, narrow band excess does not guarantee
that the emission line falling in the bandpass is Lyman $\alpha$. In the present
case, the 4970/80 passband is sensitive to Lyman $\alpha$ emission
in the redshift range [3.054,3.120], and to [OII] $\lambda 3727$ emission
in the redshift range [0.323,0.344], and even possibly to [OIII] 4959/5007 at
zero redshift. The volume probed
is larger in the [3.054,3.120] interval than in the [0.323,0.344] interval 
by a factor of between 8 and 20, depending
on cosmology; the volume probed at zero redshift is essentially negligible in comparison. 
In order to remove clear examples of ``foreground'' narrow--band
excess objects, we have cross--correlated the narrow--band catalog with
the catalog based on $U_nG{\cal R}$ photometry from the LBG survey, with the result
that 5 of the 77 NB excess objects are almost certainly foreground based upon
relatively blue $Un-G$ colors; these are indicated with ``boxes'' in Figure 2.  
Fifteen of the NB excess candidates are among the 141 Lyman break galaxy candidates
($U_n$ band ``drop--outs'') in the SSA22a field.
The remaining 57 objects in the NB catalog that satisfy the NB--excess selection criteria
are too faint in the continuum to allow meaningful limits on the presence or
absence of a Lyman continuum break (${\cal R} > 25.5$). The median
NB excess for these faint objects is ${(\rm GV -NB)_{med}=1.62}$, corresponding
to an observed emission line equivalent width of 275 \AA. As pointed out by CH98,
objects with the typical (observed) equivalent widths of $>>100$ \AA\ are much less likely 
to lie at low redshift because the implied rest--frame
equivalent widths would be highly unusual for [OII] emission, although only
spectroscopy of the candidates at high enough dispersion to resolve
potential [OII] doublets could confirm any individual case. 
We will assume in following discussion that the remaining sample of 72 line excess objects suffers from
minimal contamination by low--redshift ``interlopers'', and that the emission line
is Lyman $\alpha$ at $z \sim 3.09$. 

We have also indicated on Figure 2 objects which had been flagged previously
as candidate galaxies at $2.7 \simlt z \simlt 3.4$ 
based upon the $U_nG{\cal R}$ photometry to ${\cal R}=25.5$. 
Not all of these LBG candidates are included in the NB catalog-- about 25\%
of the LBG catalog have NB fluxes fainter than the nominal NB catalog limit of ${\rm NB=25.5}$. 
As discussed in \S 1, our follow--up spectroscopy of the LBG candidates
has yielded a total of 24 {\it known} objects in this field 
at redshifts placing Lyman $\alpha$ within the narrow--band filter. Of these
galaxies in the ``spike'',
12 would have been found using the prescribed NB color and flux criteria.
These 12 include the three most luminous narrow--band excess objects, which
includes a QSO (see Steidel \et 1998), and two very bright, highly extended
Lyman $\alpha$ nebulae, which will be discussed in detail in \S 5. 
Our spectroscopic failure rate, for observed objects,
is $\sim 25$\% for the $z\sim 3$ LBGs, but it is highly unlikely that any
spectroscopic observation of an object with an emission line exceeding 80\AA\
equivalent width would fail to yield a redshift; for this reason, the ratio
12/24 is probably a substantial over-estimate of the true fraction of
LBGs in this redshift range with strong line emission, as discussed below.  
While redshifts have been obtained for only about 40\% 
of the LBG candidates in the SSA22a field, only 3 LBG candidates
without spectroscopic redshifts have ${\rm GV-NB}$ 
large enough to make them candidates (see Figure 2). This brings to 15 the total number
of LBGs in the field which satisfy our narrow--band excess criteria. 

Four of the 24 LBGs known to lie in the spike have ${\rm GV-NB < -0.7}$, presumably
because they exhibit Lyman $\alpha$ primarily in absorption rather than in emission. 
It should therefore be possible to identify other $z \sim 3.09$ objects by the
presence of Lyman $\alpha$ {\it deficits}. In an attempt to quantify the number of objects
with significant absorption in the narrow band filter, we have plotted the positions of
${\rm GV}$--selected objects from our LBG catalog on the ${\rm GV}$ vs. ${\rm GV-NB}$ 
diagram in Figure 3. Fewer LBGs in this catalog have NB excesses than in the NB-selected
catalog because the NB flux tends to be more spatially extended than the continuum,
and isophotal apertures defined in the GV continuum band often do not include
all of the NB flux. Eighteen of the LBG candidates in this field have Lyman $\alpha$
deficits ${\rm GV-NB} < -0.7$. \footnote{
It is more difficult to quantify line equivalent width for absorption line objects, because
strong absorption usually results from a damped Lyman $\alpha$ profile with damping
wings that are significantly broader
than the width of the NB filter; nominally, $GV-NB=-0.7$ corresponds to an absorption
line that has an observed equivalent width of $\simeq 40$\AA, and the ${\rm GV-NB}$ color 
could become arbitrarily large and negative for a strong saturated damped Lyman $\alpha$ line.
However, we find that the typical color measured for objects with true equivalent
widths of $\simeq 80$\AA\ is close to ${\rm GV-NB \simeq -0.7}$}. Five  
of these have spectroscopic redshifts; only one, 
a galaxy at $z=3.238$ with such strong Lyman $\alpha$ absorption that there is still
net absorption at 4970 \AA, does not lie in the spike.  
This suggests that $4/5 \sim 80$\% of the spectroscopically unidentified objects in
the ``absorption'' region of Figure 3 also lie in the spike.

Of the 24 LBGs confirmed to lie in the spike, then, 12 satisfy our NB excess criteria, 4 satisfy
our NB deficit criteria, and 8, the remainder, have middling equivalent widths 
$\left|W_{\lambda}^0  \right| \le 20$ \AA\ that do not result in unusual ${\rm GV-NB}$ colors. 
Because our spectroscopic success rate is far higher for objects with large equivalent widths,
and because our spectroscopic completeness is only $\sim 40$\%, the proportion of all LBGs in the
spike satisfying these various criteria is unlikely to be equal to the proportions observed
in our spectroscopic sample. We estimated above that a total of 15 LBGs in the spike
satisfy our excess criteria, and that $\sim 14$ satisfy our deficit criteria. Our
40\% completeness in this field suggests that a total of $\sim 60$ (24/0.4) LBGs
lie in the spike, and so we conclude that $\sim 25$\% (15/60) of LBGs have strong
enough Lyman $\alpha$ emission to satisfy our NB excess criteria and $\sim 23\%$ (14/60)
of LBGs have strong enough Lyman $\alpha$ absorption to satisfy our NB deficit criteria. The
absolute upper limit to the fraction of LBGs satisfying our NB excess criteria is $\sim 50\%$;
this would apply if (implausibly) our spectroscopic success rate were independent of Lyman
$\alpha$ equivalent width. 

We can estimate these proportions in another way if we assume that the equivalent width
distribution of Lyman $\alpha$ in SSA22 is similar to that observed in another field
where the spectroscopic completeness is much higher, shown in Figure 4\footnote{
We should point out that the Lyman $\alpha$ equivalent width is not a well-defined
quantity in most cases, because the emission component of Lyman $\alpha$ is often
superposed on the red wing of a broad absorption feature due mostly to 
a damped Lyman $\alpha$ profile. Thus, to yield equivalent widths that are 
comparable for narrow--band photometry and spectroscopy, one would have to be very
careful to evaluate the line strength over the same range in wavelength. An additional
complication is that  the spectroscopic aperture may not contain all of the
Lyman $\alpha$ flux.  Thus, while the concept of equivalent width is well-defined
for the values estimated from photometry, it is not necessarily the same number that
would be obtained from a spectroscopic measurement.}.
Approximately 20\% of the galaxies in Figure 4 have equivalent widths that would satisfy
our NB excess criterion, and approximately 20\% have equivalent widths that would satisfy
our NB deficit criterion, consistent with the numbers estimated above.
Thus, the equivalent width
distribution inferred for {\it all} LBGs in the SSA22a field with 
${\cal R}\le 25.5$ and $z=3.090\pm0.030$ appears to be consistent with that 
shown in Figure 4; the current spectroscopic sample is biased
toward positive Lyman $\alpha$ equivalent widths, as would be expected for identifications
based on low S/N spectra. After accounting for the likely incompleteness, 
{\it only about 20-25\% of LBGs in the ground based samples are likely to be
detected on the basis of narrow--band excess}. 
The predicted total number of spike galaxies
with ${\cal R} \le 25.5$ in the observed field is $\simgt 60$, despite the fact
that only 24 have so far been spectroscopically confirmed.

\section{SPACE DENSITY AND LUMINOSITY FUNCTION OF STRONG LINE EMITTERS}     

It is of interest to compare the space density and luminosity distribution of
the objects within a known over-density at high redshift with the ``field''
galaxy luminosity function at the same redshift. Clearly, the narrow--band
catalog adds significantly to the number of known objects within the $z=3.09$
over-density, but it is difficult to make a direct comparison to the field
samples given the incompleteness with respect to continuum luminosity 
inherent in the narrow--band selection. 

A direct comparison, on the other hand, can be made to ``blank field'' surveys
for line emitters, as in CH98 and Pascarelle \et (1998); we will
focus on the former because the redshift, selection criteria,  
and depth of the observational
catalogs are more similar to our own. A direct comparison of the surface density
per unit redshift interval for the SSA22a field at $z=3.09$ and the combination
of two fields observed by CH98 (totaling 46 arc min$^{-2}$) 
to the same NB flux limit of 25.0 (AB)
yields a ratio $\Sigma_{\rm spike}/\Sigma_{\rm field} = 6.0 \pm 2.4$;
accounting for the slightly different effective volumes per unit solid angle
on the sky for the filters used by CH98 and by us yields a difference in
volume density of line emitters of $n_{\rm spike}/ n_{\rm field}=5.7\pm 2.3$
(for Einstein de Sitter). The uncertainties
are dominated by the small number statistics in the CH98 field sample.
Small corrections for differences in photometric zero point (see \S 2), 
slightly different intrinsic luminosities corresponding to fixed apparent
magnitudes between $z=3.43$ and $z=3.09$, and small differences in the
rest--frame line equivalent width limit imposed by the selection criteria are
comparatively insignificant. Thus, the enhancement in density over the
average as measured from the line emitting objects (recall that 85\% of the
emission line selected objects are {\it not} in the LBG spectroscopic sample
in Figure 1) is consistent with the enhancement inferred from the (generally)
more luminous continuum--selected objects.  

Figure 5 compares the {\it continuum} luminosity
functions of the narrow--band emitters in the ``spike'' field and
in the CH98 field sample to the LBG field luminosity
function presented by Steidel \et (1999). The continuum magnitudes
have been adjusted to correspond to observed frame ${\cal R}$ magnitudes
at $z \sim 3$ in order to facilitate the comparison between the various 
samples.  The minority of objects in both the CH98 and in the present sample
which are not significantly detected in the continuum band have been
placed in the faintest bin.  We have also included, for comparison, the narrow--band
data presented by Campos \et (1998), who suggested that the detection of 56 narrow
band emitters in the field of a CMB decrement lent support to the possibility
that a rich cluster had been detected at $z \sim 2.55$. 

The same characteristic behavior of the NB--selected samples can be
seen (modulo the difference in normalization determined above):  a very
broad range of continuum magnitudes is represented among the NB--selected
samples\footnote{The Campos \et sample is selected by continuum magnitude, rather
than by narrow--band magnitude, so that fewer faint, large equivalent width, objects
are expected}, but the samples begin to ``roll off'' at faint magnitudes
due to the nature of the selection criteria. This effect arises because
the faintest objects require very large NB excess in order
to be detected in a sample limited also by NB apparent magnitude. For example,
for continuum magnitudes of $m_{AB} =25$, the selection criterion 
${\rm GV-NB} > 0.7$ ensures that all objects whose observed
equivalent width exceeds $\sim 80$ \AA\ will be included, since all of
these objects will also have $NB < 25$. At a continuum magnitude of
26.5, on the other hand, it is only those objects which have
${\rm GV - NB} > 1.5$ that could possibly be included in a sample
limited by ${\rm NB} \le 25.0$. Thus, the relative completeness at these
two continuum magnitudes would depend directly on the intrinsic
distribution of line equivalent widths. For the purpose of illustration,
we have made an adjustment to the point centered at $m_{\rm AB}=26.0$
assuming that the true equivalent width distribution is as shown in Figure
4; in other words, the correction indicated simply accounts for the fraction
of objects which have $\rm {GV-NB} > 0.7$ but not ${GV-NB} > 1.5$, so that
the corrected luminosity function would then account for all
objects to ${\cal R} =26.5$ subject to the same rest frame equivalent
width threshold. 

From Figure 5, it is interesting to compare the $z \sim 3$ field
LBG luminosity function to that of the emission--line selected
galaxies in the $z=3.09$ ``spike''. The shapes of both are 
consistent with being the same, modulo a shift 
in normalization $\Phi^{\ast}$ by a factor of $\sim 1.5$.
Given an overall density enhancement $\simeq 6$ in the spike
(cf. Figure 1), if the fraction of galaxies
with $W_{\lambda}^0 \ge 20$\AA\ were a constant   
$f\simeq0.25$ (see \S 3) independent of luminosity, then one
would expect the emission line galaxy luminosity function in the spike to be boosted
relative to the field luminosity function by the observed factor of 1.5. 
If instead the fraction of emission line galaxies varies with luminosity, then
the continuum luminosity functions of these galaxies would also have to differ from that in
the field in such a way that the two effects conspire to cancel.
Here it should be emphasized that no photometric completeness corrections have been
made to the narrow--band selected objects; assuming an intrinsic equivalent
width distribution like Figure 4, Monte Carlo simulations indicate that
one is much more likely to include objects having true $GV-NB$ colors smaller
(i.e., with smaller line excess) than the threshold than to miss objects that should
have been included, so that in all likelihood our correction to the faintest bin
is an upper limit to the size of the appropriate upward correction. 
There is no significant evidence that fainter galaxies have
a greater probability of exhibiting strong Lyman $\alpha$ emission if the shape of
the continuum luminosity
function is the same within the ``spike'' as in the ``field''.  

Figure 5 shows that, if the fraction of line emitting
objects and the far--UV luminosity function of field galaxies remain
similar at $z \sim 2.5$ as observed at $z \sim 3$, then 
the CMB decrement field 1312+4237 observed by Campos \et (1999) 
contains only a relatively small excess of galaxies, perhaps a factor of $\sim 2$ higher
density than an average field at $z \sim 3$.
This emphasizes how difficult it is to establish 
the existence of significant over--density without a wide--area spectroscopic
survey (or at the very least, large narrow--band control fields); 
finding a large number of line emitters in a field merely reflect
the steepness of the UV luminosity function at high redshift ($\alpha = -1.6$ at $z \sim 3$ [Steidel \et 1999])
and the wide range of continuum
luminosities contained in a NB--selected sample.

\section{THE NATURE OF BRIGHT, EXTENDED LYMAN $\alpha$ EMISSION ``BLOBS''} 

Perhaps the most surprising result of the deep narrow--band imaging in the
SSA22a field is the discovery of 2 very luminous, very extended regions of
line emission, which we will descriptively call ``blobs''. These two features
are shown in gray-scale together
with the images in the GV continuum in Figure 6; their locations are indicated
on the field map in Figure 9. The properties of these
blobs are summarized in Table 2. 

As can be seen in Figure 6, the blobs appear to be at least loosely
associated with LBGs that are known to be part of the ``spike'' at
$\langle z \rangle \simeq 3.09$. Partly for this reason we consider it 
unlikely that these objects are due to emission at any other redshift; in addition, as
we will discuss below, previously existing slitlet spectra of C11 ($z=3.108$) and M14 ($z=3.091$) 
\footnote{C11 was called SSA22a-C8 in Steidel \et 1998; C15 and M14 were not known at the
time.}
confirm the basic features of the parts of the blobs that fell in the slit.
The physical nature of the Blobs can be tested further with an estimate of the emission line
equivalent width for the extended emission region; unfortunately this is
a very difficult measurement, as within the emission line region are
found several objects that are known (from their colors) {\it not} to be
at $z\simeq 3.09$ and so which must be subtracted from the continuum image
to estimate the diffuse continuum that might be associated with the 
line emission. A second problem is that at very faint flux levels in the
broad band continuum, strong line--emitting objects can be detected even
if their equivalent widths are essentially infinite; we estimate that
an object with infinite line equivalent width should have ${\rm GV-NB} \simeq 3.25$
based on the relative widths of the narrow band and effective continuum band
filters (80\AA\ versus $\simeq 1600$ \AA).  Both of the blobs have ${\rm GV-NB}$
colors (after removal of unrelated continuum sources) of $\simeq 2.5$ magnitudes;
after accounting for the emission line contamination of the broad band flux,
which in this case is $\sim $50\%, the implied equivalent width of the line emission is
$\sim$1500 \AA\ (observed frame), or a rest equivalent width of $\simeq 375$ \AA\
if the blobs are placed at $z=3.09$. Such large equivalent widths would be unprecedented
for [OII] emission at $z \sim 0.3$, and in fact are too large by at least a factor
of 2 to have been produced by stellar excitation of Lyman $\alpha$ emission (cf. Charlot \& Fall 1993;
but see below). 
The properties of the blobs summarized in Table 2 are remarkably similar 
to those of the Lyman $\alpha$ nebulae often found associated with high redshift
radio galaxies (e.g., McCarthy \et 1987; Spinrad 1989; McCarthy 1994; van Ojik \et 1997), in both Lyman $\alpha$
luminosity ($\sim 10^{44}$ ergs s$^{-1}$) and in physical extent ($\sim 100$h$^{-1}$ kpc).

Deep radio continuum observations of the SSA22 field (Yun \et 1999, private communication) 
show that any radio sources associated with the ``blobs'' must have continuum fluxes
at 1.4Ghz of $<100$ $\mu$Jy, about 3 orders of magnitude fainter than typical
powerful high redshift radio galaxies.
In further contrast to the radio galaxies, the continuum
sources that appear to be associated with the nebulosity (object C11 for Blob 1 and
M14 for Blob 2) are very faint, and in the first case not well--centered with respect to
the bulk of the Lyman $\alpha$ emission. The continuum luminosities of these objects are 2-3
magnitudes fainter than the powerful radio galaxies at comparable redshifts (the ${\cal R}$
magnitudes are 24.2 and 25.4 for C11 and M14, respectively). 
The strong line emission from high redshift radio galaxies is often thought to result from 
the interaction of jets with the ambient medium (cf. McCarthy 1994); in the case of the Blobs,
with no obvious associated radio sources, it is interesting to consider various other possibilities
for the origin of the large Lyman $\alpha$ fluxes.  

The very large line equivalent widths, as mentioned above, disfavor excitation
by a stellar radiation field; moreover, the luminosities of the known continuum sources in
the far--UV, even assuming that all photons shortward of the Lyman limit escape the
galaxy (which we already know is {\it not} the case since C11 and M14 were identified
on the basis of a strong Lyman continuum break), are at least a factor of 5 too small
to account for the line luminosities. 
The equivalent
star formation rate for an object capable of producing the observed Lyman $\alpha$ flux
associated with the Blobs
is $>100$h$_{50}^{-2}$ M$_{\sun}$ yr$^{-1}$ (the lower limit assumes Case B recombination and
no extinction of Lyman $\alpha$ photons). 
Even after nominal corrections for extinction
using the UV colors, the extinction--corrected star formation rates
are $\sim 17$ M$_{\sun}$ yr$^{-1}$ for both C11 and M14 ($H_0=50$, $q_0=0.5$) (Adelberger \et 2000). 

Another possibility, consistent
with the large equivalent widths and the absence of a plausible exciting source, would
be that the gas seen in emission is not in photoionization equilibrium, but instead is
gas cooling from much higher temperatures, as in a cooling flow. 
The recombination rate implied by the luminosity of 
$\sim 10^{44}$ ergs s$^{-1}$ (here we will assume that $\Omega_m=0.3$,
$\Omega_{\Lambda}=0.7$, and $h=0.7$, as the calculations are highly
cosmology dependent)  implies $\sim 10^{62}$ recombinations per year, or a mass
deposition rate of $\sim 10^{5}/n$ M$_{\sun}$ yr$^{-1}$, where $n$ is the number
of recombinations per baryon (cf., e.g.,  Heckman \et 1989). The most massive
dark matter halo that is likely to be present at $z \sim 3.1$ assuming a CDM--like
model with the above cosmology is $\sim 10^{13}$ M$_{\sun}$; the implied
cooling would then cool all of the baryons (assuming 15\% baryon fraction) in such
a halo from the virial temperature on a timescale of $\sim 10^{7}n$ years. This timescale
is long enough that it is perhaps plausible to observe two similar objects within a fairly
small, but unusually over--dense, volume. The associated cooling timescale would be much shorter
for an Einstein-de Sitter cosmology because of the smaller baryon fraction ($\sim 5$\%) and
much smaller plausible dark matter halo mass ($\sim 10^{12}$ M$_{\sun}$).
Observations of cooling flow clusters at low redshift generally show H$\alpha$
emission that implies multiple recombinations per baryon in order to be consistent
with the cooling rate inferred from the X--ray gas (Kent \& Sargent 1979;
Heckman \et 1989), i.e., $n >>1$. The most massive cooling flow clusters at low redshift produce
H$\alpha$ luminosities of $\sim 10^{43}$ ergs s$^{-1}$ (Edge \et 1994), which could
plausibly have Lyman $\alpha$ luminosities approaching those of the Blobs. 
It would be difficult to imagine that the Blobs are associated with the mass
scales ($\sim 10^{15}$ M$_{\sun}$) that produce substantial cooling flows at lower
redshift, for any popular cosmological model; nevertheless, it remains possible
that the emission from the Blobs is somehow related to the cooling flow phenomenon,
after accounting for the possibility that gas densities could be much higher at $z\sim 3$
than at lower redshift. 

Still another possibility is that the continuum photons responsible for ionizing the
Lyman $\alpha$ emitting gas are obscured from our view, and yet escape in other directions
sufficiently to produce large Lyman $\alpha$ fluxes which are partially scattered into
our line of sight. To test this possibility, in 1999 May we obtained deep $K_s$ images
of the Blobs using the Keck I 10m telescope and the Near Infrared Camera (NIRC) (Matthews
and Soifer 1994);  these images are shown superposed on the line emission images of
the Blobs in Figure 7. 
While we detect the two known LBGs associated with the
Blobs (object C11 and M14 for Blob 1 and Blob 2, respectively), 
the optical/IR colors are not unusual with respect to the bulk of LBGs, 
although object M14, with ${\cal R}-K_s=3.7$, is in the reddest quartile of LBGs at
comparable redshift (cf. Shapley \et 1999). Possibly more interesting is
the $K_s$ band source near the center of Blob 1 (see Figure 7) that is coincident
with the brightest ``knot'' of Lyman $\alpha$ emission after that associated with
C11. This object is not obviously present in our line--free continuum bands (${\cal R}$, I) in the optical; it has
$K_s=21.2$ and has a color of ${\cal R}-K_s > 4.5$. If the object is associated with
the Lyman $\alpha$ nebulosity (as opposed to being a chance projection of a red object), then
it is possible that it is heavily obscured and therefore much more luminous than it appears. Assuming that its intrinsic
colors are similar to that of the nearby object C11 (${\cal R}-K_s=2.6$, which is
consistent with zero extinction for our LBG spectral template at $z =3.1$), it could have
an intrinsic luminosity more than 9 times brighter than that of C11, assuming a Calzetti (1997) attenuation
curve,  implying a star formation rate of 
$\simgt 160$M$_{\sun}$ yr$^{-1}$ ($q_0=0.5$, $h=0.5$).
In any case, if the Lyman $\alpha$ emission is excited
by star formation, with most of the UV radiation obscured from our line of sight, 
the large Lyman $\alpha$ luminosity would necessarily place the blobs in the range of star formation rates
that approaches those detectable at sub--mm wavelengths. 
Maps of a portion of the SSA22a field at 850$\mu$m have recently been
obtained by Barger \et (1999); unfortunately, only Blob 2 falls within the region mapped
(and it falls near the edge, where the sensitivity is reduced significantly). The 
corresponding upper limit ($3\sigma$) is $<7.2$mJy, which is not very constraining as it
corresponds to an upper limit on the equivalent star formation rate of $\sim 1000$ 
M$_{\sun}$ yr$^{-1}$. 
In this context it is interesting
to ask whether luminous Lyman $\alpha$ nebulae might accompany some of the known
SCUBA sources. It is conceivable that complicated geometrical and radiative transfer
effects could conspire to allow substantial Lyman $\alpha$ fluxes at the same time
that the continuum sources are heavily obscured (cf. Neufeld 1991).  

Also shown on Figure 7 are the slit orientations for the original spectra of C11 and M14 obtained
with LRIS in 1996 October and 1997 September, respectively, as part of our LBG survey. The
observations were acquired using slit masks and the 300 line/mm grating, which provides
a spectral resolution of $\sim 12.5$ \AA. 
Both spectra were obtained using 1\secpoint4 slitlets which happen to fall
across parts of the blobs; the total exposure times were 9600s and 6400s for
C11 and M14, respectively. 
In Figure 8 we present spectrograms showing the
spatial extent and the velocity field of the parts of the blobs which were of high enough
surface brightness to be significantly detected, and which happened to fall on the slits\footnote{These slit PAs
were dictated by the preferred orientation of the multi--object slit masks without
{\it a priori} knowledge of the existence of the Blobs}. Blob 2 exhibits
a particularly interesting velocity field, with differences of $\sim 2000$ \kms
evident within a few arc seconds of the continuum object M14. The amplitude of the velocity
field near M14 is again reminiscent of high redshift radio galaxies (e.g. van Ojik \et 1997). 
In the case of C11, less diffuse emission is detected, but the knot of Lyman $\alpha$ emission
$\sim $7\arcs\ east along the slit is clearly detected spectroscopically. These
detections leave little doubt that the emission from Blobs 1 and 2 is at $z \sim 3.1$; neither
spectrum shows any significant detection of lines other than that identified with Lyman $\alpha$. It
would clearly be of interest to obtain spectra at different position angles through the
Blobs, preferably at higher spectral resolution and with greater sensitivity to low surface
brightness emission.

\section{MAPPING THE $z=3.09$ STRUCTURE}

One of the primary motivations for obtaining the deep narrow--band images
of the volume containing the LBG over-density was to map out the galaxy
distribution in more detail.
Figure 9 shows the locations of objects with either
secure spectroscopic identifications placing them within the
redshift--space over-density, or with narrow--band excesses
or deficits which plausibly place them in the same volume. 
The total number of (independent) objects within the volume is 162,
after augmenting the $NB < 25$ narrow--band excess sample with the fainter sample
of narrow band excess objects having $25.0 \le NB \le 25.5$ and $GV-NB > 1.0$ discussed
in \S 3. 
The redshift interval sampled by the narrow--band imaging is [3.054,3.120],
so that the total dimensions of the mapped region are as given in \S 2,
approximately 8${\rm h}^{-1}$ by 8${\rm h}^{-1}$ Mpc on the plane of the sky
and $\simeq 23{\rm h}^{-1}$ Mpc in depth for an Einstein-de Sitter universe. 

While the visual impression is that the distribution of galaxies within
the volume has regions of apparent local
over-density and under-density relative to the average, a $w(\theta)$ analysis
shows that there is no evidence for significant angular clustering above
that expected for a random distribution. This does not mean that
there are not real substructures-- only that the structure is
not enhanced over that expected for objects laid down at random within the volume. 
In Figure 10, we show a surface density map smoothed with a Gaussian with $\sigma=30$
arc seconds (i.e., FWHM $\sim 1{\rm h}^{-1}$ Mpc co--moving), 
and contoured in units of $\delta \equiv {\rho-{\bar \rho} \over {\bar\rho}}$.
Here an over-density of 1 corresponds to a region that is $\sim 2 $ times denser
than the average density within the ``spike'' volume, or $\sim 12$ times denser
than the average density in the general field (cf. \S 4).  The three highest peaks 
in the local density are significant at the 99.8\% confidence level or better, based on
a counts in cells analysis of many random catalogs using 
circular cells of radius equal to the smoothing scale used
to produce Figure 10; again, this number of 3$\sigma$ peaks in the local
density is not unexpected even for a random catalog, as found above. 

On the other hand, the most striking aspect of Figure 10
is that two of the objects of special interest, Blob 1
and the $z=3.083$ QSO (see Steidel \et 1998) appear to lie near the centers of two
of the most prominent over-dense regions. Blob 2, on the other hand, lies in a region that
is near the mean spike density, and there is a third ``peak'' on the western side of the field 
with a compact (projected) structure but no known object of unusual interest, aside from 3 spectroscopically
confirmed LBGs (see Figure 9), which have redshifts of 3.076, 3.097, and 3.103. 
It is of course possible that projection effects could conspire to produce apparent
over-densities on the plane of the sky, given that the dimension of the volume along the line of
sight is $\sim 3$ times larger than the transverse dimensions. With only 24 spectroscopic redshifts
out of 162 objects, assigning individual objects to real substructures is not really possible;
in fact, even with redshift information there would be a great deal of ambiguity, given that
the redshift measurements accessible in the far--UV (including, especially, the Lyman $\alpha$
line) can exhibit large offsets relative to systemic velocities (e.g., Steidel \et 1998,
Pettini \et 1998, 1999). For this reason, it is hard to argue that extensive spectroscopic effort
to obtain redshifts for a greater fraction of the objects in Figure 9 would yield unambiguous clues
to (e.g.) the dynamical state of this proto--cluster region. 

Generically, one might expect sub--structure within a region whose
overall density enhancement suggests that it will evolve into a rich cluster of galaxies
by the present epoch. In this sense, Figures 9 and 10 offer a first glimpse of a large
number of objects in such a region at $z \sim 3$, and the appearance might be compared
in a qualitative way with expectations from simulations (see, e.g., Governato \et 1998,
Wechsler \et 1998).
We are at present limited by the relatively small transverse dimensions of
the field; expectations from simulations would be that galaxies are forming along sheets
and filaments, and that large over--dense regions occur at the intersections 
of these topological features. Our small two-dimensional projection is rather limited in its
usefulness for such important comparisons; much larger fields could now be covered using
state of the art CCD mosaics and a rather substantial investment in observing time.
Using narrow--band imaging to map out the extent and topology of large structures
could be a very important observational test in the future.  

While we have found that the projected distribution of objects within the over-dense
volume at $\langle z \rangle = 3.09$ is not significantly different from random,
the coincidence of the locations of Blob 1 and the QSO 
(called SSA22a-D13 by Steidel \et 1998) relative to the projected 3$\sigma$ peaks
is highly significant. 
If QSOs are indeed the sites of rare, massive dark matter halos at
high redshift (cf. Haehnelt \& Rees 1993), it is perhaps not surprising  
that a generally over-dense region would contain a QSO, and that
the QSO would be located in a region of local density enhancement. We do not really 
know if objects like the Blobs are unusual or not, as there have been very few observations
made in which they could have been detected. The fact that one of them is found within
an apparent local density enhancement in a region that is already significantly over-dense
with respect to an average place in the universe at $z \sim 3$ suggests that perhaps
they too are unusual phenomena that may not be found in the general ``field''. The possible
association of Blob 1 with either a cooling flow--like phenomenon, an obscured but
very luminous continuum source, or even a ``dead'' radio galaxy, 
would be qualitatively consistent with finding
Blobs in ``special'' locations. It may well be that such large Lyman $\alpha$ nebulae
are the denizens of the highest density regions at high redshift, with some fraction
of them associated with active radio sources.  At this point, given the lack of
adequate ``control'' fields,  or further sampling of other over-dense regions 
at high redshift, one can only speculate.

\section{SUMMARY AND DISCUSSION}

1. We have obtained a very deep narrow--band image of an 8\minpoint7 by 8\minpoint9 field
centered on a significantly over-dense region in the ``SSA22'' field that was
discovered during the course of a spectroscopic survey for photometrically selected Lyman break
galaxies at $z \sim 3$. The narrow--band images allow the detection of objects
with strong Lyman $\alpha$ emission ($W_{\lambda}^0 > 20$ \AA) 
in a volume centered at $z =3.09\pm 0.03$. There is a total
of 72 NB--excess objects in the primary catalog, of which 13 had been previously flagged
as Lyman break galaxy candidates, and 11 had previous spectroscopic redshifts placing them within
the targeted redshift range. The remainder of the NB--excess objects are fainter in the
continuum than the limit of the Lyman break technique from the ground.   
Comparison with published narrow--band results to comparable sensitivities and at
similar redshifts indicates that the SSA22a field is similarly over-dense in faint objects
as it is in bright ones, with a density enhancement of about $5.7 \pm 2.3$, as compared
to $6.0 \pm 1.2$ estimated from the generally brighter LBG sample. 
The uncertainties in this comparison
are currently dominated by the very small ``field'' samples of NB--selected high
redshift galaxies. 

2. Narrow--band photometry targeting the Lyman $\alpha$ emission line 
is very efficient at detecting large numbers of high
redshift galaxies, most of which are significantly fainter than in the
spectroscopic LBG samples. 
However, it provides a sample that is 
subject to complicated selection effects with respect to more physically
relevant quantities such as far--UV continuum luminosity. The current
sample, together with the spectroscopic results from SSA22 and other LBG
survey fields and the far--UV LBG luminosity function presented by Steidel \et (1999), 
suggests that about
20-25\% of all LBGs to continuum luminosities of ${\cal R} \sim 27$ are detected
given the typical sensitivity of narrow--band observations. This fraction does
not appear to depend significantly on continuum luminosity. Thus, 
emission--line selected samples are probably very incomplete at all redshifts
and at all luminosities. One should bear this in mind whenever high redshift galaxy
surveys depend on the strength of the Lyman $\alpha$ line. 

3. We have discovered two objects, which we call ``Blobs'', which are very
extended $(\simgt 15$ \arcs), diffuse, and luminous 
($L({\rm Ly} \alpha ) \sim 10^{44}$ ergs s$^{-1}$)  
Lyman $\alpha$ nebulae, with many 
properties similar to the giant Lyman $\alpha$ nebulae associated with
high redshift radio galaxies. However, these Blobs are associated with relatively
faint objects of the LBG class and have no detected radio emission at the
100 $\mu$Jy level. We have considered several explanations for the Blobs,
with no firm conclusion emerging. Possible explanations include the presence of  
embedded ionizing sources that are obscured from our view (perhaps supported
by deep 2.15 $\mu$m continuum observations), but whose
UV radiation field may escape in other directions 
sufficiently to excite the extended Lyman $\alpha$ nebulae, or that the nebulae
are high redshift analogs of massive cooling flows observed in clusters at low redshift. 
While these are the
first such nebulae (of which we are aware) to be discovered without associated
radio sources, there is a dearth of sensitive narrow--band observations covering large fields
at high redshift; as a result, it is unclear whether or not this phenomenon is unusual. 

4. Given the peculiar nature of narrow--band selection for Lyman $\alpha$, 
which generally depends both
on equivalent width and line flux of a line which is notoriously difficult to quantify,
perhaps the most useful application of the technique is to ``map out''
particular structures at high redshift. While we have shown that only 
about 20-25\% of the objects (at a given far--UV continuum flux) are
likely to be detected in this way, the technique is sensitive to a much broader
range of continuum luminosity than, e.g., the LBG surveys with typical
ground--based sensitivity. As a result, 
the surface density of emission line objects,
particularly within over-densities such as the one discussed in this paper, 
can be very high, and can allow detailed topological analysis of the galaxy distribution
for structures at very high redshift. We have explored this application for the
SSA22 ``spike'' at $z=3.09$; there is a total of 162 objects that are either known
or suspected to be associated with the spike in the small field, after
combining the NB--excess catalogs, a small catalog of ``NB--deficit'' objects which are
also LBG candidates, and the existing spectroscopic LBG catalog (which has 23 galaxies
and 1 QSO associated with the ``spike''). The resulting map yields the best 
high redshift ``snapshot'' of a region that 
is in all probability destined to be a rich cluster by the present epoch.  
A crude analysis of
the projected surface density distribution yields an interesting result: while
the overall distribution of objects is consistent with random, of the
3 highest local peaks ($\ge 3 \sigma$) in the surface density,  one is centered
on a $z=3.083$ QSO discovered in the original spectroscopic survey, and another
is centered on one of the two luminous Lyman $\alpha$ ``Blobs''. 
While we refrain from making too much of this result at present, given the many
uncertainties involved, it does suggest that the ``Blobs'' may represent a different
class of rare object which, like QSOs, are preferentially found in rich environments
at high redshift.

\bigskip
\bigskip

We are indebted to Amy Barger and Min Yun for communicating sub--mm and radio
results for the SSA22 field prior to publication, and to David Frayer and
Steve Rawlings for illuminating conversations. 
CCS acknowledges support from the U.S. National Science Foundation through
grant AST 95-96229, and from the David and Lucile Packard Foundation. 
MG has been supported through grant HF-01071.01-94A from the Space Telescope
Science Institute, which is operated by the Association of Universities for
Research in Astronomy, Inc. under NASA contract NAS 5-26555.

\bigskip
\newpage
\newpage
\begin{deluxetable}{lccccl}
\tablewidth{0pc}
\scriptsize
\tablecaption{Summary of Photometric Catalogs}
\tablehead{
\colhead{Catalog}  & \colhead{Detection} & \colhead{Type of} & 
\colhead{Number} &
\colhead{Number of ``Spike''}  & \colhead{Photometric Selection} \nl
\colhead{}  & \colhead{Isophotes} & \colhead{NB Selection} & 
\colhead{Selected} &
\colhead{Objects Selected\tablenotemark{a}} & \colhead{Criteria}
} 
\startdata
NB-selected & NB & Excess & 72\tablenotemark{b}  & 12 & NB $\leq 25.0$ ; GV$-$NB $\geq 0.7$\tablenotemark{c} \nl
 & &  & & & \nl
LBG GV-selected & GV & Deficit & 18  & 4 & GV$-$NB $\leq -0.7$ ; LBG $U_nG{\cal R}$ criterion\tablenotemark{d}\nl
\enddata

\tablenotetext{a}{Galaxies satisfying the NB excess/deficit
photometric criteria, and which are also spectroscopically confirmed members 
of the $\langle z \rangle= 3.09$ structure.}

\tablenotetext{b}{While 77 objects in the NB-selected catalog satisfy the 
NB $\leq 25.0$ ; GV$-$NB $\geq 0.7$ NB-excess criteria, 5 have $Un - G$ colors
which indicate that they are foreground objects, and therefore should not be included in the present analysis.}

\tablenotetext{c}{A fainter sample of objects with $25.0 \leq$ NB$ \leq 25.5$, GV$-$NB $\geq 1.0$
was added to the original sample of NB $\leq 25.0$ objects, for the purposes of
mapping structure at $z=3.09$ (\S 6). The combined sample numbers 136 objects
selected by NB excess.}

\tablenotetext{d}{LBG galaxies satisfy ${\cal R}\leq 25.5$, (G$-{\cal R}$) $\leq 1.2$, and 
(G$-{\cal R}$) + 1.0 $\leq$ ($Un -$G) $\leq$ (G$-{\cal R}$) + 1.5  (Steidel {\it et al.} 1998)}
\end{deluxetable}

\begin{deluxetable}{lccccccc}
\tablewidth{0pc}
\scriptsize
\tablecaption{Properties of the Lyman $\alpha$ ``Blobs''}
\tablehead{
\colhead{} & \colhead{Coordinates}  & \colhead{Angular Size} & \colhead{GV\tablenotemark{a}} & 
\colhead{NB\tablenotemark{a}} & \colhead{$W_{\lambda}$(Ly $\alpha$)\tablenotemark{b}} &
\colhead{F(Ly $\alpha$)} & \colhead{L(Ly $\alpha$)\tablenotemark{c}} \nl
\colhead{} & \colhead{(J2000)} & \colhead{(arc sec)} & \colhead{(AB mag)} & \colhead{(AB mag)} 
& \colhead{(\AA)} & \colhead{(ergs s$^{-1}$ cm$^{-2}$)} & \colhead{(ergs $s^{-1}$)} 
} 
\startdata
Blob 1 & 22 17 25.7  +00 12 49.6 & 17 & 23.52 & 21.02 & $\simgt$370 & $1.4\times10^{-15}$ & $1.0\times10^{44}$ \nl  
Blob 2 & 22 17 39.0  +00 13 30.1 & 15 & 23.55 & 21.14 & $\simgt370$ & $1.2\times10^{-15}$ & $9.0\times10^{43}$
\nl  
\enddata
\tablenotetext{a}{Corrected for contamination by unrelated foreground objects}
\tablenotetext{b}{Limit on the rest equivalent width, assuming that the line is Lyman
$\alpha$ at $z=3.09$}
\tablenotetext{c}{For $H_0=50$ \kms Mpc$^{-1}$, $q_0=0.5$}
\end{deluxetable}
\newpage
\medskip
\begin{figure}
\figurenum{1}
\plotone{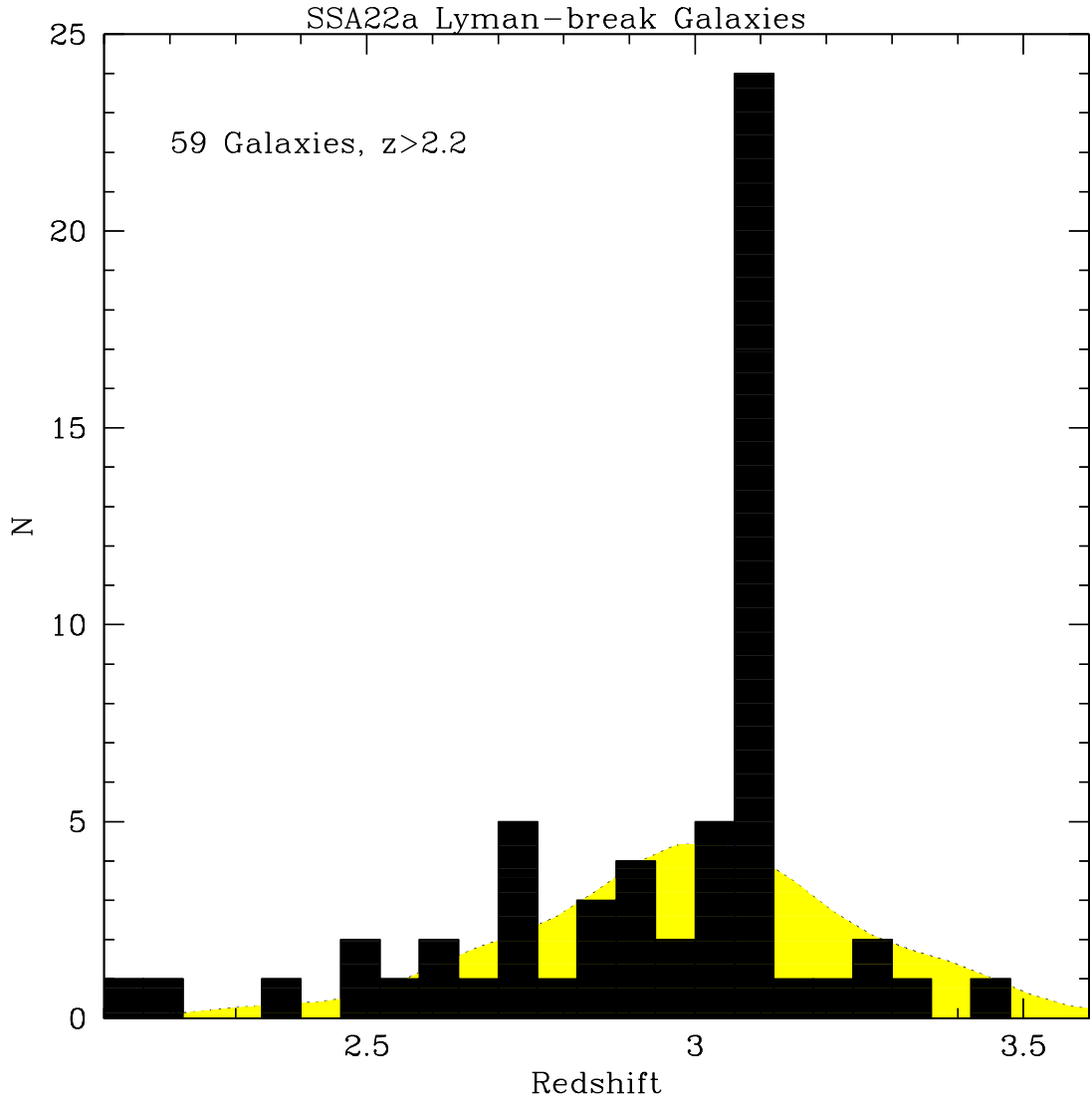}
\caption{Histogram of Lyman break galaxies with spectroscopic redshifts, from
the region also imaged in the narrow--band 4970/80 filter. The histogram has
been binned such that the prominent ``spike'' is centered on the
median redshift of the galaxies in the structure, with a bin width corresponding
to the redshift range probed by an 80\AA\ filter at $z \sim 3$. Note that the
density enhancement relative to the overall survey selection function (the light, smooth
histogram) is a factor of $\sim 6\pm1.2$ (cf. Steidel \et 1998).}
\end{figure}
\begin{figure}
\figurenum{2}
\plotone{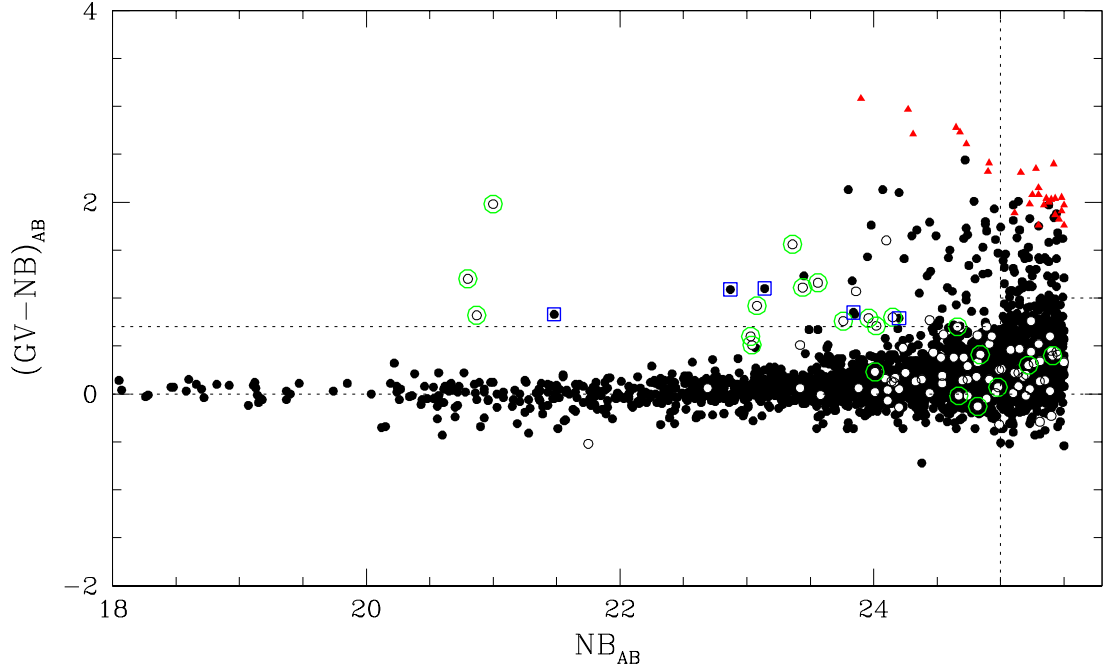}
\caption{Diagram of objects detected in the NB--selected catalog to an
apparent magnitude limit of NB$=25.5$. The dotted horizontal line
for $NB \le 25.0$ (the adopted limit for the principle
catalog of line emitters lies to the left of the vertical dotted line) 
is drawn at $(GV-NB)=0.7$; objects above this line
have emission lines with observed equivalent widths of $\sim 80$\AA\ or greater.
Triangles indicate objects that are not detected in the continuum at better
than the 2$\sigma$ level. 
Objects which were flagged as candidate Lyman break galaxies at $z \sim 3$ 
are indicated with skeletal (white) dots;  
objects which have spectroscopic redshifts placing them in the interval [3.054,3.120]
are circled. The 5 objects above the horizontal dotted line which have
broad--band colors placing them at $z << 3$ are indicated with boxes. 
A fainter catalog of line emitters is drawn from objects having $(GV-NB)>1$
and $25.0 \le NB \le 25.5$.   
}
\end{figure}
\begin{figure}
\figurenum{3}
\plotone{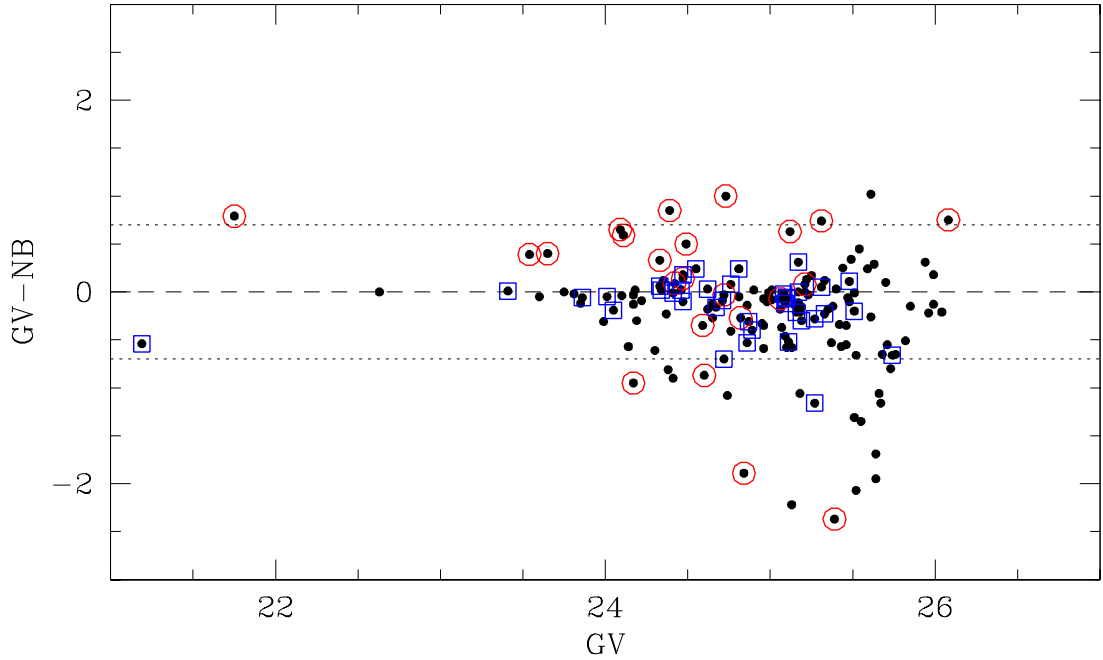}
\caption{ Narrow band colors for all objects from the GV--selected catalog that
also satisfy our LBG selection criteria in $U_nG{\cal R}$ color space. The
known objects in the $z=3.09$ structure are indicated with circles. Objects
with spectroscopic redshifts for which Lyman $\alpha$ does {\it not} fall
within the 4970/80 narrow band filter are indicated with a square. Note
that there are fewer objects in this catalog which would be found as narrow--band
excess objects, and that there is a large number of objects with
negative colors (i.e., effective absorption in the NB) that remain unidentified
in our $U_nG{\cal R}$--selected catalog.}
\end{figure}
\begin{figure}
\figurenum{4}
\plotone{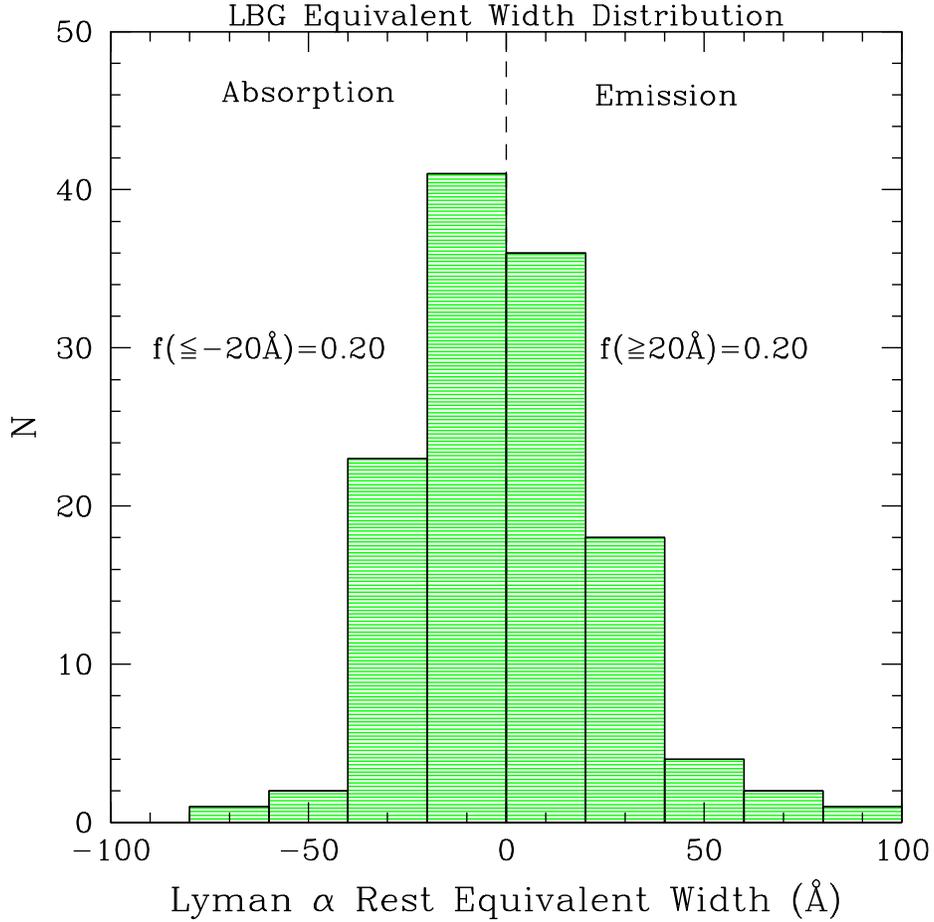}
\caption{The distribution of spectroscopically measured Lyman $\alpha$
rest equivalent widths from one of the Lyman break galaxy survey fields
(the ``Westphal'' field-- see Giavalisco \et 1998)
in which the spectroscopy is 92\% complete for objects having
${\cal R}\le 25.0$. Note that the median Lyman $\alpha$ rest equivalent
width is near zero, and that about 20\% of the objects exhibit Lyman
$\alpha$ emission that could be detected using narrow band imaging, with
the same fraction exhibiting absorption strong enough to be detected
as narrow--band deficits (due to absorption). 
The statistics of LBGs in the SSA22a field at $z=3.090$
field are entirely consistent with these fractions (see text).}  
\end{figure}
\begin{figure}
\figurenum{5}
\plotone{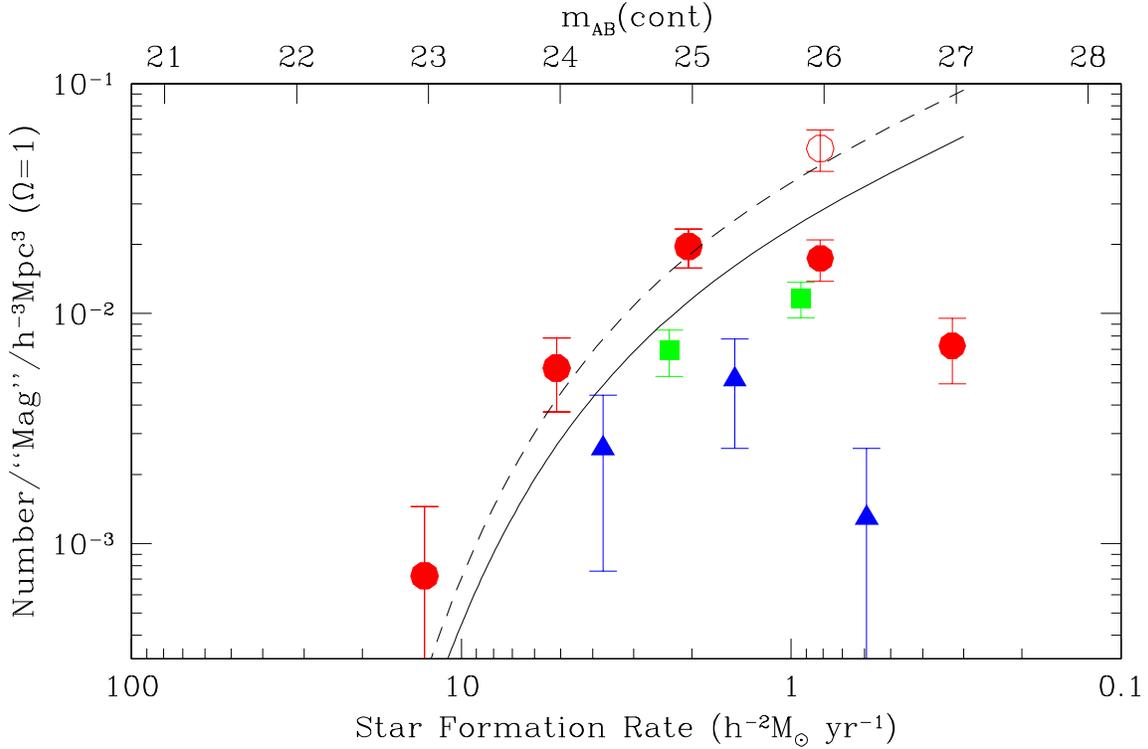}
\caption{Comparison of the continuum luminosity function of Lyman $\alpha$ emission--selected
objects (circles with error bars) to the luminosity function for 
$z \sim 3$ LBGs from Steidel \et (1999) (solid
line) after
excising low z objects and the QSO. The triangles are the data from a combination of 
the two $z=3.4$ ``blank field'' narrow--band pointings of Cowie \& Hu (1998). Note that luminosity
function of the ``field'' narrow band emitters is well below that of the LBGs 
at a similar redshift, and is a factor of $\sim 6$ times below that of the
$z=3.090$ ``spike'' data. The dashed curve is the LBG luminosity function with
$\Phi^{*}$ increased by a factor of 1.5. The unfilled point at $m_{AB}=26$
represents a correction that assumes that the true equivalent width
distribution for the line emitters is as shown in Figure 4; see \S 4 for discussion.
The narrow--band data of Campos \et (1999) ($z \sim 2.55$) are shown with squares; note
that this field represents only a small over-density if the field LF and the
fraction with strong emission lines are similar at $z \sim 3$ and $z \sim 2.5$. 
 }
\end{figure}
\begin{figure}
\figurenum{6}
\plotone{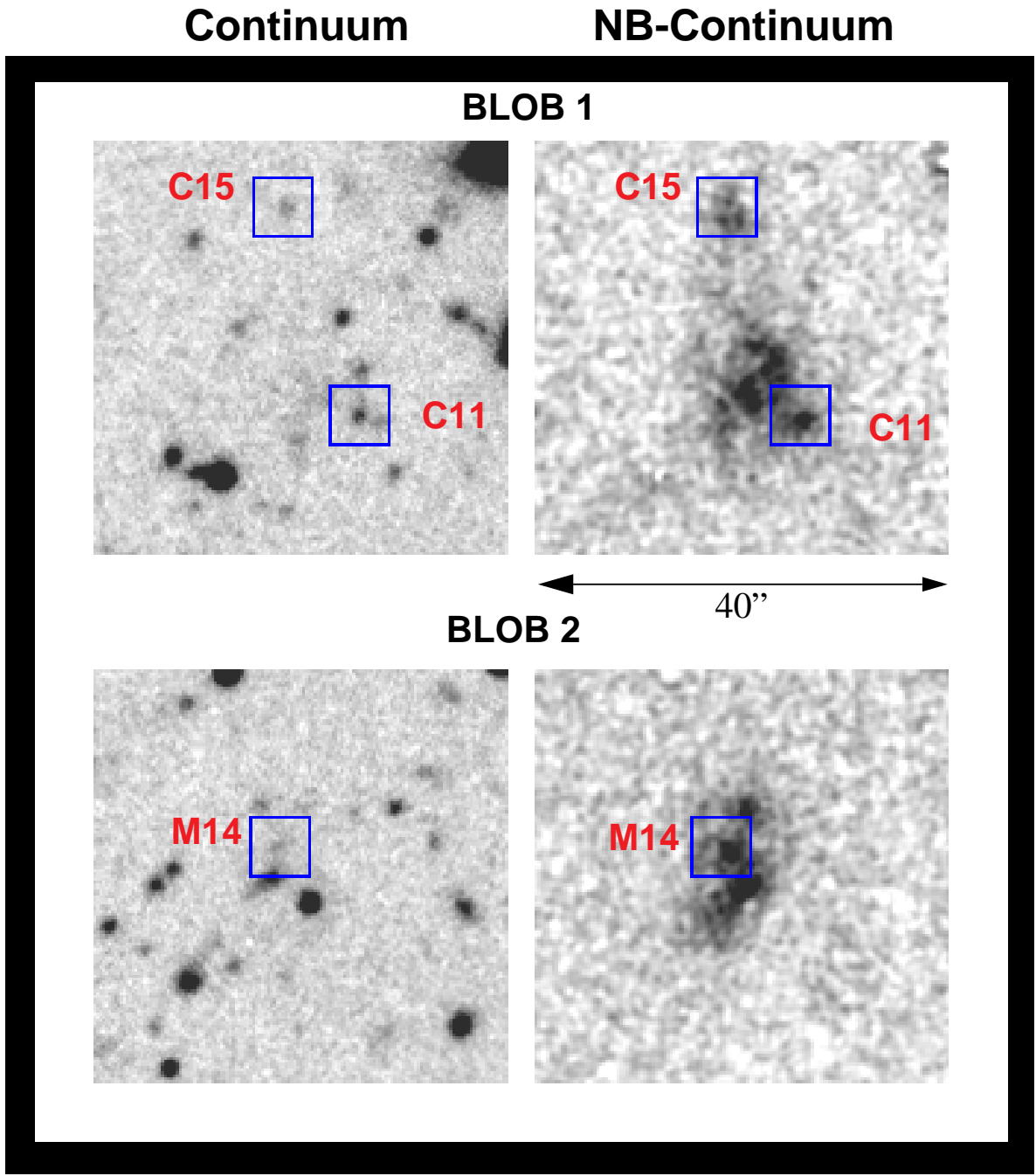}
\figcaption{ Portions of the GV continuum and continuum--subtracted NB images showing
the two prominent ``Blobs''. The boxes indicate LBGs with spectroscopic redshifts
placing them within the $z=3.090$ over-density. The properties of the Blobs
are summarized in Table 2.
}
\end{figure}
\begin{figure}
\figurenum{7}
\plottwo{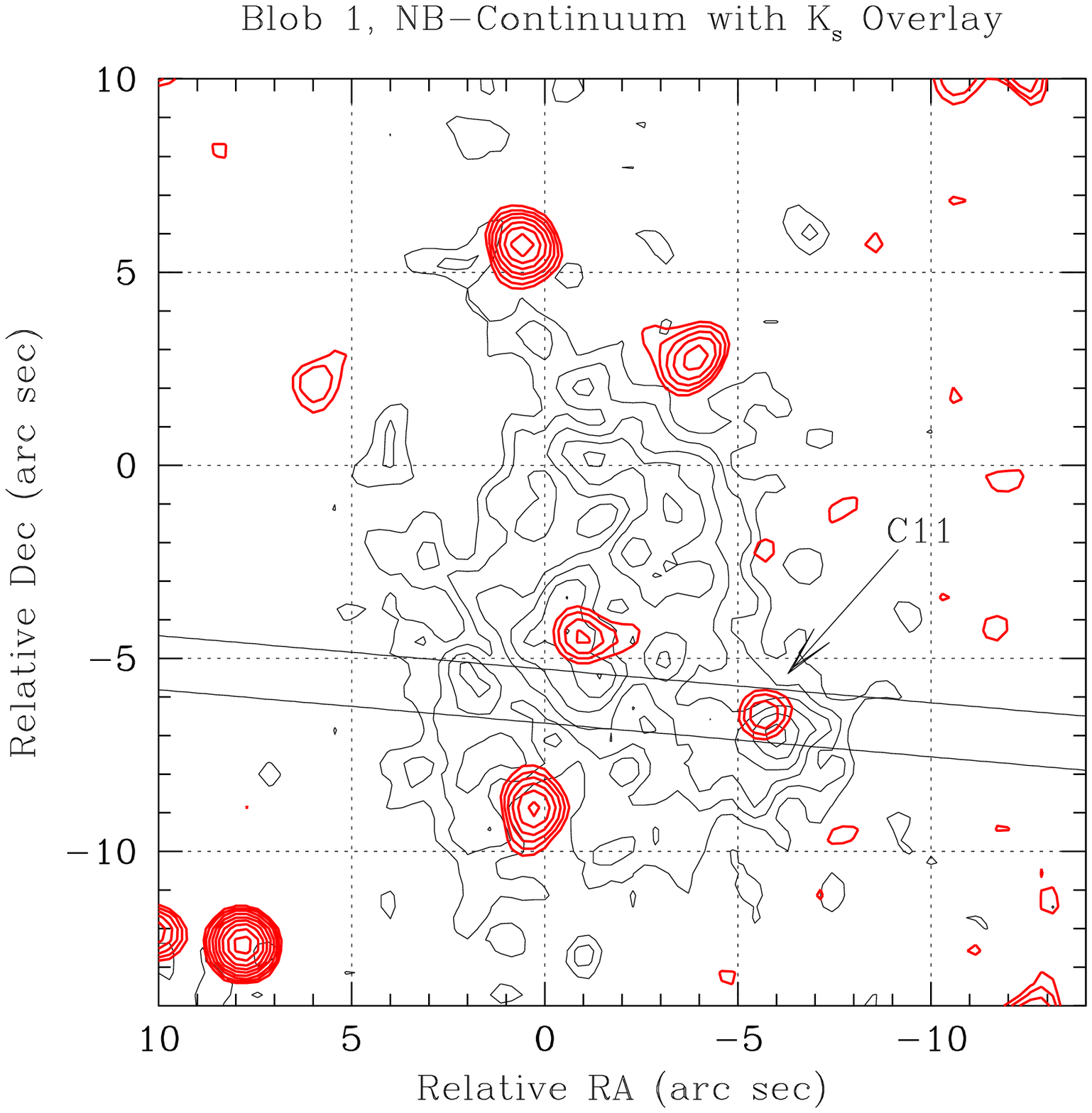}{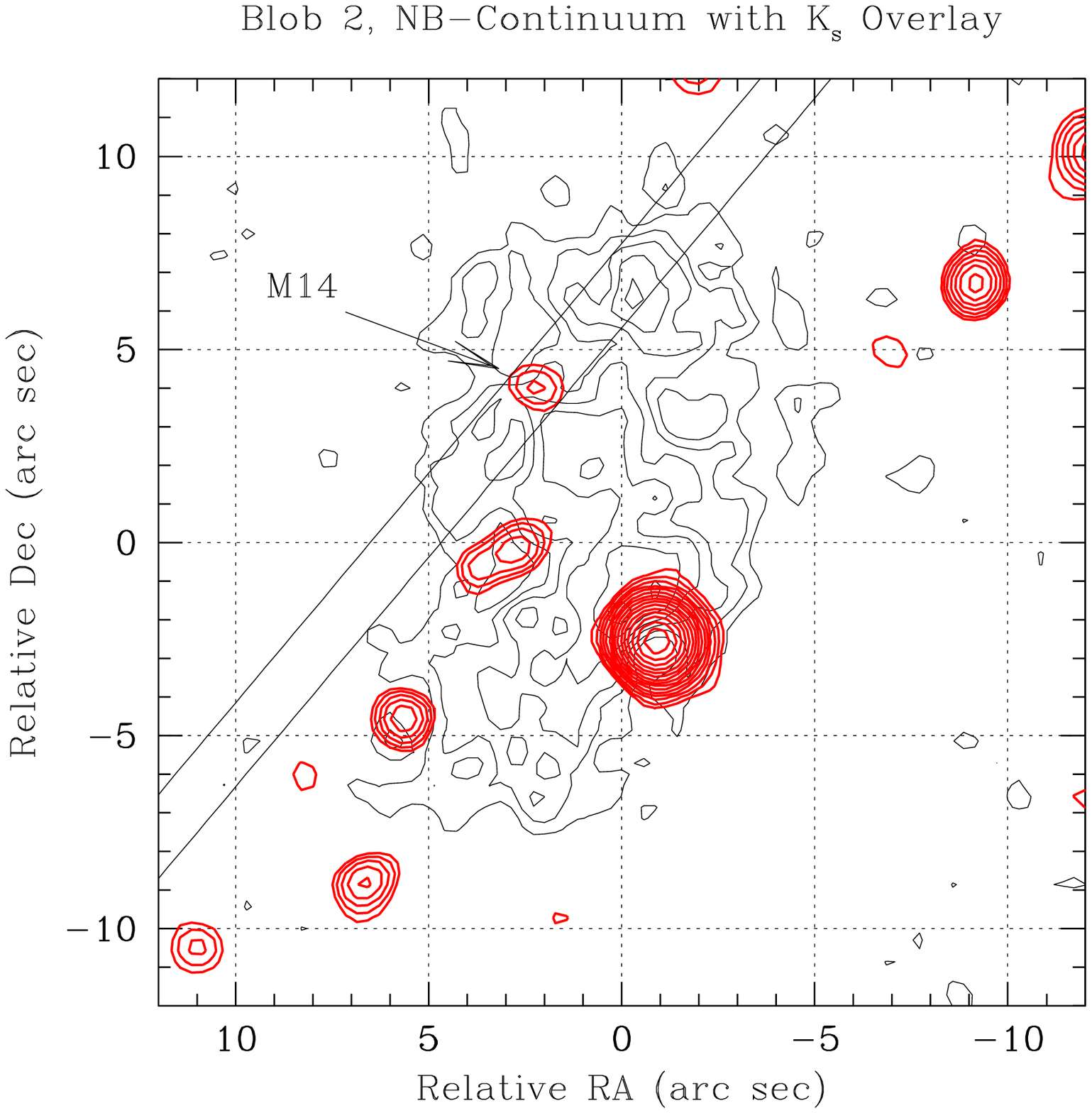}
\caption{ Contour maps of the Blobs (light contours) in NB-Continuum light, with
Keck $K_s$ images superposed; N is up, E to the left. 
Note that there is a $K_s$ band source that is spatially
coincident with the central knot in Blob 1. It is estimated that this object has
a color of ${\cal R}-K_s \simgt 4.5$, making it about 1.9 mag. redder in that color
than the known LBG, C11 (see Figure 6 for comparison).} 
\end{figure}
\begin{figure}
\figurenum{8}
\plottwo{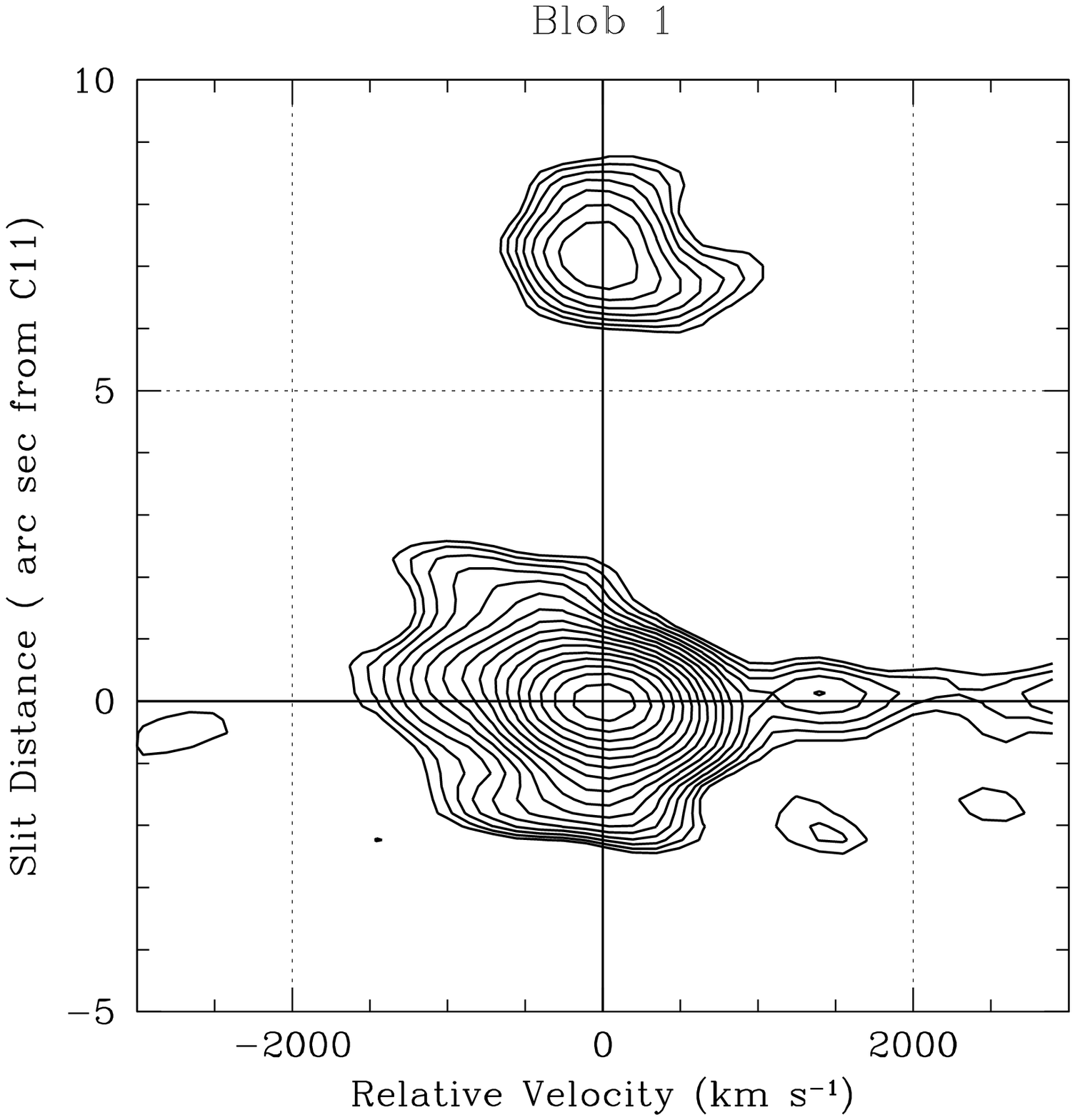}{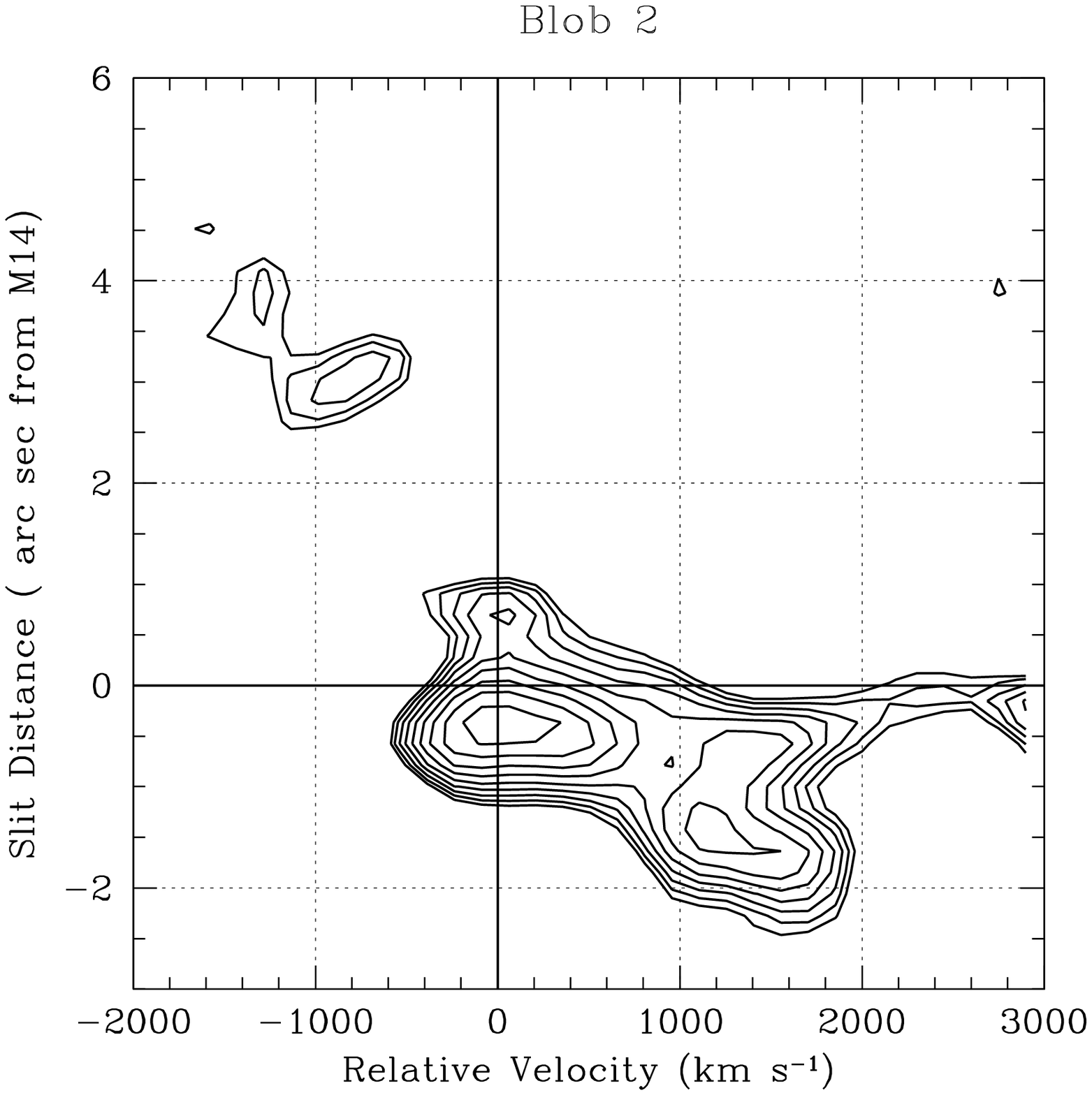}
\caption{Spectra in the vicinity of Lyman $\alpha$ for C11 (left, with E toward positive
slit positions) and for M14 (right, with SE toward positive slit positions). Note the
apparent velocity shear of the Lyman $\alpha$ emission near M14, and the clear detection of
the emission knot 7\arcs\ E of C11.}
\end{figure}
\begin{figure}
\figurenum{9}
\plotone{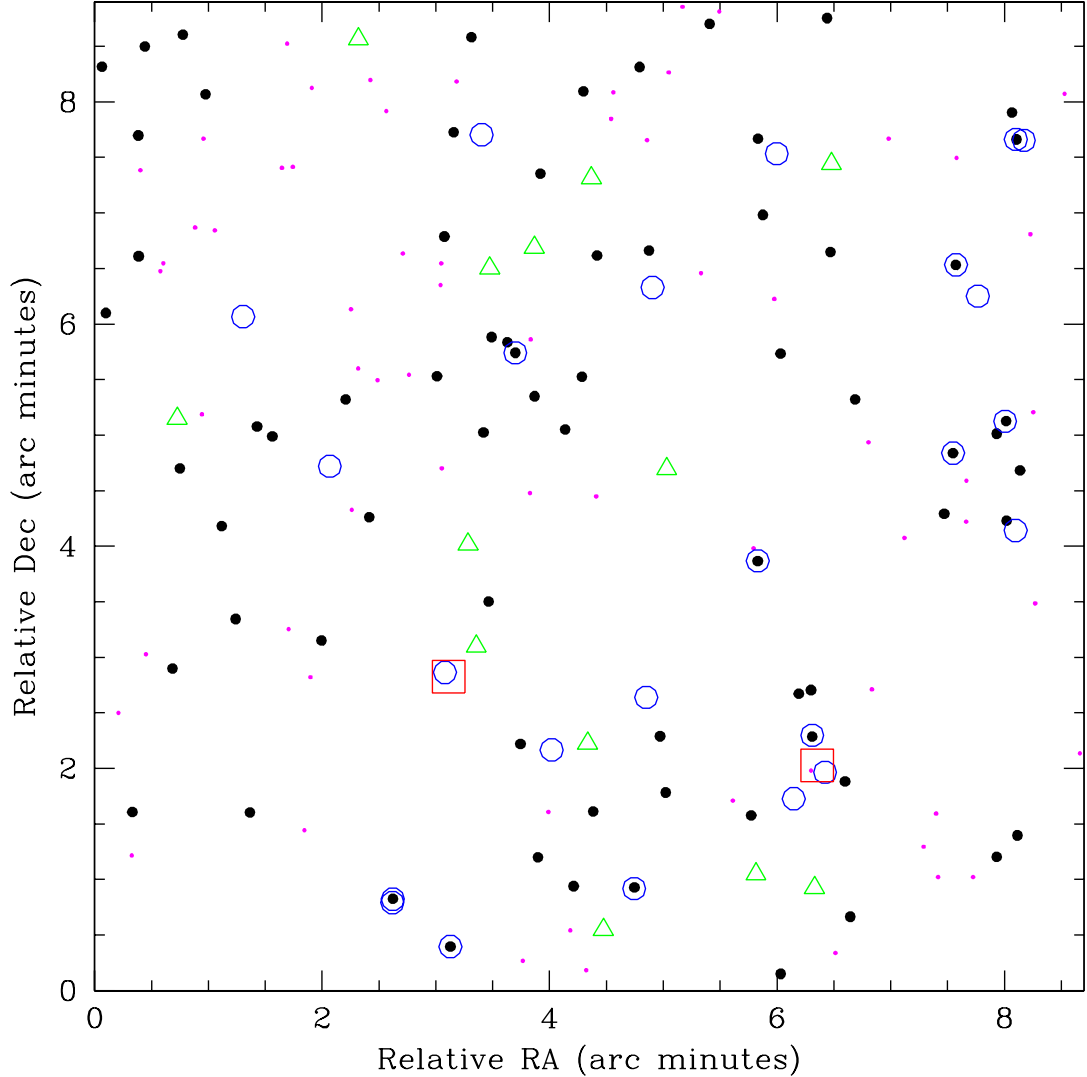}
\caption{A map (North is up, East to the left)
of the known and suspected ``spike'' objects in the
SSA22a field. Open circles represent LBGs with spectroscopic redshifts in the
range $3.06 \le z \le 3.12$ (24 objects), larger dots represent narrow--band
excess objects from the $NB=25.0$ sample (72 objects, excluding the 5 known to be
in the foreground, which are not plotted), 
smaller solid dots have $25.0 < NB \le 25.5$ and $GV-NB>1.0$
(64 objects), open triangles are LBG candidates with significant
narrow--band deficits but without spectroscopic redshifts, 
and the large open squares are the two giant line emitting ``Blobs''. There is
a total of 162 objects in the volume that are either known or likely
to be associated with the $z=3.090$ over-density. 
 }
\end{figure}
\begin{figure}
\figurenum{10}
\plotone{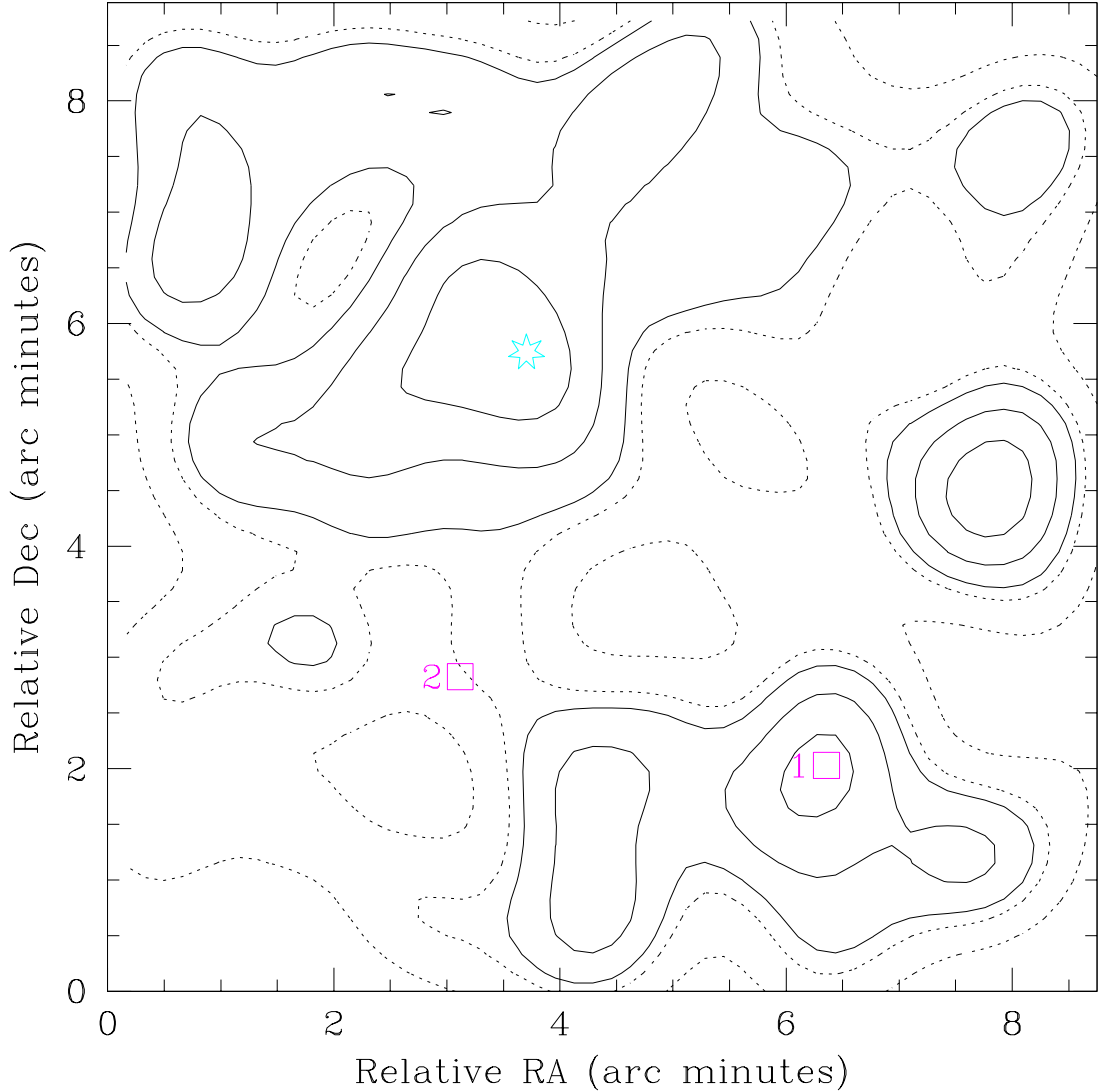}
\caption{Smoothed surface density map of $z \simeq 3.09$ objects. The map was generated with
a Gaussian smoothing kernel of $\sigma=30$ arc seconds, or FWHM$\sim 1$h$^{-1}$ Mpc (co--moving)
at $z \sim 3.1$. The contours indicate deviations from the mean density; the solid contours
represent $\delta \equiv {\rho-{\bar \rho} \over {\bar\rho} }= 0.1$,0.5,1,1.5, while the dotted contours are 
$\delta= -0.1$,$-0.5$,$-1$. The positions of the QSO (star symbol) and Blobs 1 and 2
(boxes) are indicated. Note that the QSO and Blob 1 are very close to the centers
of 2 of the 3 most prominent local over-densities; these features are all significant
at greater than the 3$\sigma$ significance level.}     
\end{figure}
\end{document}